\documentclass[aps, prb,reprint, superscriptaddress, floatfix,longbibliography]{revtex4-2}

\usepackage{lipsum}  
\usepackage{graphicx}
\usepackage[cmyk,dvipsnames]{xcolor}
\usepackage{amsbsy}
\usepackage{amsmath}
\usepackage{amssymb}
\usepackage{amsfonts}
\usepackage{amsopn}
\usepackage{bigints}
\usepackage[utf8]{inputenc}
\usepackage[T1]{fontenc}
\usepackage{placeins}
\usepackage{float}
\usepackage{blindtext}
\usepackage{diagbox}
\usepackage[normalem]{ulem}
\usepackage{multirow}
\usepackage{bm}
\usepackage{physics}
\usepackage{braket}
\usepackage{hyperref}
\hypersetup{
    colorlinks,
    linkcolor={blue},
    citecolor={blue},
    urlcolor={blue}
}
\setcitestyle{titles}
\usepackage{booktabs}

\usepackage{rotating}

\usepackage{algorithm}
\usepackage{algpseudocode}
\algnewcommand{\LineComment}[1]{\State // #1}
\algrenewcommand\textproc{\texttt}

\newcommand{\defcommand}[3][1]{
    \providecommand{#2}{}
    \renewcommand*{#2}[#1]{#3}
}

\defcommand[1]{\mat}
{
    \boldsymbol{#1}
}

\definecolor{pacificb}{HTML}{1CA9C9}

\newcommand{\kiel}{Institute of Theoretical Physics and Astrophysics, Christian-Albrechts-University, 24118 Kiel, Germany}
\newcommand{\iceland}{Science Institute and Faculty of Physical Sciences, University of Iceland, VR-III, 107 Reykjav\'{i}k, Iceland}
\newcommand{\kalmar}{Department of Mathematics and Physics, Linnaeus University, SE-39231 Kalmar, Sweden}

\let\oldvec\vec
\renewcommand{\vec}[1]{\oldvec{#1}\,}

\begin{document}

\title{Network of localized magnetic textures revealed using a saddle-point search framework}

\author{Hendrik Schrautzer}
\affiliation{\iceland}

\author{Tim Drevelow}
\affiliation{\kiel}

\author{Hannes J\'{o}nsson}
\affiliation{\iceland}

\author{Pavel F. Bessarab}
\email[Corresponding author: ]{pavel.bessarab@lnu.se}
\affiliation{\iceland}
\affiliation{\kalmar}

\begin{abstract}
A computational framework is presented for the exploration of
energy surfaces of magnetic systems via systematic identification of first-order saddle points that 
connect local energy minima representing possible states of the system. 
The saddle points characterize the mechanism of transitions between states and, together with the energy of the local minima,
provide estimates of the transition rates within harmonic transition state theory or Kramers/Langer theory.
The calculations involve four stages: (1) the symmetry of a given energy-minimum configuration is identified and used to define subsystems whose eigenmodes provide relevant deformation directions.
(2) These modes are then used to guide the system toward the neighborhoods of the various saddle points on the energy rim surrounding the given energy minimum. 
(3) Next, the geodesic minimum mode following method is employed to efficiently converge onto the saddle points.
(4) Finally, the identified saddle points are embedded into the network describing the connectivity of the states of the system. 
This methodology is applied in a study of the metastable textures of two-dimensional chiral magnets described by a lattice Hamiltonian, where a hierarchy of transition mechanisms for the nucleation, annihilation, and rearrangement of 
various localized magnetic textures are identified, such as skyrmions, antiskyrmions, skyrmion bags, chiral droplet, etc. 
Transitions corresponding to homotopies that preserve the topological charge as well as transformations that change it are identified. 
By scaling the interaction parameters, the distinct behavior of these two classes is obtained as the continuum limit is approached. 
For some transitions, the topological charge changes during the transition even though it is the same for the initial and final states.
\end{abstract}



\maketitle

\section{Introduction}
\label{sec:intro}
In recent years, the field of topological magnetism has advanced significantly with the discovery of various localized magnetic textures in 2D and 3D, extending beyond magnetic skyrmions~\cite{goebel2021}.
Magnetic systems capable of hosting multiple structures simultaneously~\cite{rozsa2017,rybakov2015,rybakov2019,kuchkin2020,kuchkin2022} are of particular interest, as they hold great promise for technological applications and offer a potential for interesting phenomena driven by the interactions and transformations between different magnetic textures.

The system’s ability to accommodate several stable states signifies the presence of multiple minima on its energy surface.
How the system traverses between these minima due to thermal activation can be studied using harmonic transition state theory (HTST)~\cite{wigner1938,vineyard1957,bessarab2012} or Kramers/Langer theory~\cite{kramers1940,langer1969,fiedler2012,bocquet2024}.
The identification of first-order saddle points (SPs) on the energy surface is the key part of this analysis, with each SP defining the mechanism and the rate of possible transitions between the states.

In the context of topological magnetism, the energy minima correspond to magnetic textures that can be sorted into classes according to their topological charge. In 2D systems, the topological charge is the skyrmion number~\cite{nagaosa2013}
\begin{align}
    Q=\frac{1}{4\pi}\int \left[\vec{m}\cdot\left(\partial_x \vec{m}\times\partial_y \vec{m}\right)\right]\dd{r}^2~,
    \label{eq:topo_charge}
\end{align}
where $\vec{m}(\vec{r})$ denotes the magnetization. For two magnetization textures with the same topological charge $Q$, it is possible to define a homotopy -- a continuous transformation between them~\cite{schwarz_1994,kuchkin2022}; such a transformation is not unique. A continuous transformation between magnetic textures with distinct $Q$ is not possible as it inevitably leads to the formation of discontinuities of the magnetization~\cite{schwarz_1994,rybakov2019}. 
When a magnetic configuration is represented by an arrangement of magnetic moments $\vec{m}_i$ on discrete lattice sites, both homotopic and non-homotopic transformations \footnote{Understood in the context of a continuum interpolation of the lattice configuration.} are possible, and both correspond to overcoming finite energy barriers
~\cite{lobanov_2016,heil2019,kuchkin2022,kuchkin_2022droplet}.

The existence of multiple transition mechanisms -- both between textures with the same topological charge and between states with different $Q$ -- motivates the search for SPs connected to a given initial-state minimum, without prior knowledge of the final states. 
Such a single initial point
problem differs from the double initial point problem where both the initial and final states are specified and the SP can be located by finding the minimum energy path (MEP) connecting them~\cite{jonsson1998,henkelman2000NEB,mathiesen2019,bessarab2015}.


Algorithms designed for single initial point problems~\cite{
mousseau1998,henkelman1999,peters2017} 
typically involve two stages. In the first, 
an {\it escape stage}, the system is 
driven out of
the convex region of the initial state minimum. In the second, 
{\it a convergence} stage, the system is 
brought to the nearest
first-order SP
on the energy surface.
Implementations of the convergence stage rely on the observation that first-order SPs 
can be reached by 
ascent 
along the
direction parallel to the eigenvector of the Hessian corresponding to the lowest eigenvalue -- the minimum mode. Maximizing the energy along the minimum mode, while minimizing it along all orthogonal directions, ultimately leads to convergence to a
first order SP.

SP search algorithms have been extended to magnetic systems where special care must be taken to account for the curvature of the configuration space arising from the constraint on the length of the magnetic moments. These developments have mostly focused on the convergence stage~\cite{muller2018,bocquet2023,sallermann2023,schrautzer2025}, where the main challenge is the evaluation of the minimum mode. In particular, recent implementation of the geodesic minimum mode following (GMMF) method enables an efficient determination of the minimum mode without explicit evaluation of the Hessian, making it applicable to large magnetic systems, including those with long-range interactions~\cite{schrautzer2025}.



An energy minimum can be connected to multiple SPs, each representing a particular transition mechanism. To sample these properly, several SP searches need to be performed, with the convergence stage in each search starting from sufficiently distinct configurations to ensure that different valleys of the energy surface are explored and different SPs are reached. These entry points for the convergence stage are generated during the escape stage. Therefore, the escape stage is a crucial part of the SP searches, directly affecting the completeness of sampling of SPs surrounding a given minimum. An effective strategy for the escape stage should yield a diverse and unbiased ensemble of entry configurations, ultimately enabling systematic identification of distinct SPs.


Previous implementations of the escape stage have typically pursued an eigenmode following strategy~\cite{henkelman1999, muller2018,bocquet2023,schrautzer2025}. The system is 
initially displaced randomly from the energy minimum, and then
iteratively pushed uphill in energy along a selected Hessian eigenvector, recomputed at each iteration, until an escape criterion is met.
Typically, only a subset of the degrees of freedom of the system is perturbed in the initial displacement~\cite{henkelman2001,Pedersen2011,EON}.
Different initial displacements can lead to convergence on different SPs even if the minimum mode is followed in each case.
A sufficient diversity in the initial displacements results in efficient sampling of the SPs.
Choosing the initial displacements to be widely distributed on a hypersphere has been shown to increase the number of different SPs identified in the convergence stage while the overall number of attempts is reduced~\cite{plasencia2017}. 
Recently, multi-objective genetic algorithms~\cite{deb2002} have been applied to learn from previous SP search attempts and improve the diversity and quality of initial displacements~\cite{xu2026}.
Following
low-energy eigenmodes is a natural approach, but it has some limitations. 
Many important SPs -- such as those corresponding to localized symmetry-breaking transformations -- may involve deformations of only a small part of a magnetic texture. Introducing a tendency toward such a SP during the escape stage may require following several eigenmodes sequentially~\cite{schrautzer2025}, making the approach less systematic. 



In this article, a strategy for the escape stage in single initial point SP searches in magnetic systems is presented and applied to transitions between various magnetic textures. 
The strategy is based on the selective excitation of subsystems of magnetic textures. Such localized excitations break the symmetry of the initial configuration and generate tendencies toward SPs associated with deformations of the textures. Together with symmetry-informed sampling of the subsystems, efficient convergence to SPs, and their subsequent classification, the approach constitutes a framework for the systematic exploration of the energy surfaces of magnetic systems. Its efficiency is demonstrated for a 2D chiral magnet, where a complex network of energy minima and SPs is revealed. The SPs represent mechanisms of transitions between localized magnetic textures, and both homotopic and non-homotopic transformations are found in the network. A hierarchy of energy barriers between stable states is identified, and its behavior under scaling toward the continuous magnetization limit is analyzed. 
The article is organized as follows. Sec.~\ref{sec:model} introduces the model of a chiral magnet and describes the properties of localized magnetic textures in a 2D system. Sec.~\ref{sec:method} presents the SP search framework, including the pre-processing, escape, convergence, and postprocessing stages. Sec.~\ref{sec:results} describes the results of characterization of the energy surface of a two-dimensional chiral magnet within the developed framework. 
Sec.~\ref{sec:conclusion} concludes the article with a summary and discussion of the findings. 

Three appendices provide additional details: (A) the correspondence between the lattice Hamiltonian and the continuum micromagnetic functional; (B) the effect of escape-stage thresholds on the outcome of the subsequent convergence stage; and (C) an additional test of how strongly the energy barriers depend on the shape of the magnetic texture.


\section{Model}
\label{sec:model}
\subsection{Hamiltonian}
\begin{figure}
    \centering
    \includegraphics[width=1.0\linewidth]{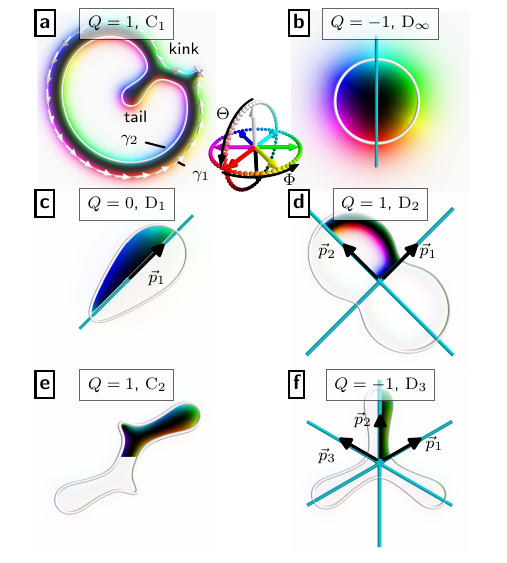}
    \caption{Examples of magnetic textures with different topological charges $Q$ and symmetries, described by the spin point groups $\mathrm{C}_n$ and $\mathrm{D}_n$. The textures correspond to energy minima for $h=0.623$ and $L_D=40a$. The $m_z=0$ contours are indicated by light-gray lines. The inset illustrates the hue–saturation–lightness color scheme, in which the hue encodes the azimuthal angle $\Phi$ of the magnetic moments, the lightness represents the polar angle $\Theta$, and the saturation is fixed at $1.0$. 
    \textbf{a}: Texture featuring a chiral kink, a tail and two nested closed contours $\gamma_1$ and $\gamma_2$. The magnetic moments along the outer contour are shown with the white arrows.
    \textbf{b}-\textbf{f}: Skyrmion with axial symmetry (\textbf{b}), chiral droplet (\textbf{c}), skyrmion bag with two inner contours (\textbf{d}), two-tailed texture with two chiral kinks (\textbf{e}), and three-tailed texture (\textbf{f}). The mirror axes are shown in cyan, with their orientations denoted by $\vec{p}_j$. The fundamental domains of the textures are highlighted, while the remaining parts are grayed out.
    }
    \label{fig:introduction_textures}
\end{figure}

The isolated textures in 2D magnets~\cite{rybakov2019,kuchkin2020,kuchkin2023} (see Fig.~\ref{fig:introduction_textures} for examples) 
are modeled on a square lattice, where the magnetic configuration $\bm{m} = (\vec{m}_1, \dots, \vec{m}_N)$ is described by $N$ unit vectors $\vec{m}_i$ specifying the directions of the magnetic moments at the lattice sites, each of magnitude $\mu$. 
The corresponding lattice Hamiltonian includes contributions from the Heisenberg exchange, the Dzyaloshinskii-Moriya (DM) interaction, and the Zeeman interaction:
\begin{equation}
\label{eq:hamiltonian}
    \begin{split}
        E = &- \sum\limits_{n}\frac{J_n}{2}\sum\limits_{\braket{ij}_n}\vec{m}_i\cdot\vec{m}_j\\
        &-\sum\limits_{n}\frac{D_n}{2}\sum\limits_{\braket{ij}_n}\vec{d}_{ij}\cdot(\vec{m}_i\times\vec{m}_j)\\
    &-\mu\sum\limits_{i}\vec{B}\cdot\vec{m}_i~.
    \end{split}
\end{equation}
Here, the outer sum in the first two terms runs over the neighbor shells $n = 1, 2, \ldots$, and the inner sum includes all sites $i, j$ separated by the distance corresponding to the $n$-th neighbor shell, with the strengths of the Heisenberg exchange and DMI characterized by $J_n$ and $D_n$, respectively.
The unit DM vector $\vec{d}_{ij}$ is given by the following formula: 
\begin{equation}
\label{eq:dm}
    \vec{d}_{ij}=R_{\beta}\vec{r}_{ij}/|\vec{r}_{ij}|,
\end{equation}
where $\vec{r}_{ij}$ connects sites $i$ and $j$ and $R_\beta\in\mathbb{R}^{3\times 3}$ describes the right-handed rotation around the normal to the system plane by angle $\beta$.
For $\beta=0$, this corresponds to Bloch DM interaction~\cite{yu2010,yu2011,yu2015}, while $\beta=\pi/2$ describes N\'eel-type chiral modulations~\cite{romming2013,kezsmarki2015,romming2015}. 
The external magnetic field 
$\vec{B}$ is perpendicular to the system plane and periodic boundary conditions within the system plane are used.

A 2D chiral magnet can also be described within the continuum model of magnetization, in which the energy of the system 
is given by the following micromagnetic functional
\begin{equation}
\label{eq:hamiltonian_micromag}
\begin{split}
    \mathcal{E}=\bigintssss\Bigg[&\frac{\mathcal{J}}{2}\sum_{\alpha=x,y}(\partial_\alpha\vec{m})^2+\mathcal{D}R_\beta\vec{m}\cdot\vec{\nabla}\times R_\beta\vec{m}\\
    &-M_S(\vec{B}\cdot\vec{m}-B)\Bigg]\dd{r}^2+\mathcal{E}^{\text{FM}}.
\end{split}    
\end{equation}
Here, $\vec{m}=\vec{m}(\vec{r})$ is the unit vector field of magnetization, 
%
$\mathcal{E}^{\text{FM}}$ is the energy of the ferromagnetic (FM) state, $M_S$ is the saturation magnetization, and $\mathcal{J}$ and $\mathcal{D}$ 
are the exchange and DM interaction parameters of the continuum model, respectively. 

A long-wavelength expansion of the lattice Hamiltonian yields a micromagnetic energy functional (see Appendix~\ref{app:continuous_discrete}) that reproduces the functional in Eq.~\eqref{eq:hamiltonian_micromag} to second order in the spatial derivatives of the magnetization. This establishes a mapping between the parameters of the continuum and lattice models~\cite{Rybakov2022}.
Higher-order contributions can be eliminated by an appropriate choice of the lattice parameters, thereby improving the correspondence between the models. In particular, in this work a lattice model with three neighbor shells is employed, for which the interaction parameters are chosen such that the two descriptions coincide up to fourth order in the spatial derivatives of the magnetization~\cite{donahue1997,buhrandt2013} (see Table~\ref{tab:modelparameters}). 


In this work, the equilibrium period of chiral modulations in the ground state, $L_D=2\pi\mathcal{J}/\mathcal{D}$, is varied, and the unit of energy is $\mathcal{J}$. Within the lattice model, increasing $L_D$ reduces discretization effects and approaches the continuum regime, in which the characteristic length scale $L_D$ is large compared to the nearest-neighbor distance $a$. The magnetic field is parametrized by the dimensionless quantity $h$, such that $B=hB_D$, where $B_D=\mathcal{D}^2/(M_S\mathcal{J})$ is the saturation field. The lattice Hamiltonian parameters used in this work are listed in Table~\ref{tab:modelparameters}.
%
%
%
%


\begin{table}[t]
\centering
\caption{
Parameters of the lattice Hamiltonian in Eq.~\eqref{eq:hamiltonian}, chosen such that the continuum limit reproduces the micromagnetic functional in Eq.~\eqref{eq:hamiltonian_micromag} up to fourth order in the spatial derivatives of the magnetization.}
\begin{ruledtabular}
\begin{tabular}{cccccccc}
$J_1$ & $J_2$ & $J_3$ & $D_1$ & $D_2$ & $D_3$\\
\midrule
$\frac{4}{3}\mathcal{J}$ & $0$ & $-\frac{1}{12}\mathcal{J}$ & $\frac{8\pi}{3L_D}a\mathcal{J}$ & $0$ & $-\frac{2 \pi}{3L_D}a\mathcal{J}$ 
\\
\end{tabular}
\end{ruledtabular}
\label{tab:modelparameters}
\end{table}

 
\subsection{Isolated magnetic textures}
Magnetic textures emerging as metastable states in two-dimensional chiral magnets, whose energy is described by Eqs.~(\ref{eq:hamiltonian}) and (\ref{eq:hamiltonian_micromag}), can take a wide variety of forms. Nevertheless, any isolated texture can be described in terms of three fundamental building blocks:
\begin{itemize}
    \item \textbf{Closed contours:} Lines in space separating regions of opposite out-of-plane magnetization. The sense of magnetization rotation along a contour is determined by the chirality of the system set by the DM interaction. Closed contours can be nested within one another, with the outermost contour defining the boundary of the localized texture~\cite{rybakov2019}.
    \item \textbf{Chiral kinks:} Localized twists of the magnetization along a contour where the sense of rotation reverses relative to the intrinsic chirality of the system~\cite{kuchkin2020}.
    \item \textbf{Tails:} Local distortions or extensions of a contour that modify the overall shape of the texture but, unlike kinks, do not alter the sense of magnetization rotation~\cite{kuchkin2023}.
\end{itemize}
By combining these elementary building blocks in various ways, a surprisingly large diversity of configurations can be obtained. Whether a given configuration corresponds to a true energy minimum ultimately depends on the specific parameters of the Hamiltonian. Figure~\ref{fig:introduction_textures}a shows an exemplary texture involving all types of the constituent elements: two closed contours, a chiral kink, and a tail. This texture corresponds to a minimum of Eq.~(\ref{eq:hamiltonian}) for a magnetic field of $h = 0.623$ and $L_D = 40a$. 

All magnetic textures predicted by the 2D chiral magnet model fall into distinct homotopy classes defined by their topological charge $Q$ [see Eq.~(\ref{eq:topo_charge})], which can be expressed as the sum of the winding numbers of the magnetization along all closed contours~\cite{kuchkin2020}:
\begin{equation}
\label{eq:topo_charge1}
    Q=\sum\limits_{i}\Omega_i,
\end{equation}
where the winding number $\Omega_i$ associated with contour $\gamma_i$ is defined as
\begin{align}
\label{eq:winding}
    \Omega_i=\frac{1}{2\pi}\oint\limits_{\gamma_i}\vec{\nabla}\Phi(\vec{r})\cdot\dd{\vec{r}}~,    
\end{align}
with $\Phi(\vec{r})$ being the azimuthal angle of the magnetization. The orientation of each contour is defined such that, when the contour is traversed, the region with $\Theta > \pi/2$ lies on its left.

This representation illustrates how nested contours, chiral kinks, and tails contribute to the total topological charge. Each contour without kinks contributes a winding of $\pm 1$, depending on the sense of magnetization rotation relative to the contour orientation. For the metastable configurations considered here, each chiral kink adds one positive winding to its host contour. 
Note, models with strong easy-axis anisotropy can also support metastable textures with negative chiral kinks~\cite{kuchkin2020}.
Tails affect only the geometric shape of the texture and do not modify its topological charge~\cite{kuchkin2023}. For example, an isolated skyrmion contains a single contour without kinks, yielding $Q = -1$, while more complex multi-contour, multi-kink configurations can have $Q = 0, \pm1, \pm2, \ldots$, depending on the number and arrangement of the constituent elements [see Fig.~\ref{fig:introduction_textures}].

The topological charge is a key property that governs many physical phenomena, such as the topological Hall effect~\cite{wang2022} and the skyrmion Hall effect~\cite{yang2024}. However, $Q$ alone is insufficient to uniquely specify the structure of a texture. Distinct configurations may share the same topological charge yet differ in morphology and symmetry [see Fig.~\ref{fig:introduction_textures}a and Fig.~\ref{fig:introduction_textures}d]. Moreover, the symmetry of a texture has a direct impact on its dynamics~\cite{kuchkin2021,kuchkin2025}.


To further distinguish localized textures, their symmetry properties can be analyzed. In this work, we identify texture symmetries with respect to $n$-fold rotations ($\mathrm{C}_n$) and $n$-fold rotations combined with reflections ($\mathrm{D}_n$). This analysis is used during the preprocessing stage of the developed framework to define the \textit{fundamental domain} of each magnetic texture -- loosely speaking, the smallest region containing all unique physical information about the texture without repetition (Sec.~\ref{ssec:method_preprocess}).
The strategy of the escape stage is based on excitations of various subsystems centered on the points within the fundamental domain, as described in Sec.~\ref{ssec:method_escape}. Examples of fundamental domains for several magnetization textures are shown in Figs.~\ref{fig:introduction_textures}c–f. 

The wide diversity of localized textures, including distinct states sharing the same topological charge, makes the 2D chiral magnet model an excellent platform for studying physical realizations of homotopies -- continuous transformations between textures of the same $Q$. Textures belonging to the same homotopy class can always be continuously transformed into one another, and infinitely many such homotopies exist. In contrast, transformations between textures of different homotopy classes are impossible without forming discontinuities in the magnetization~\cite{rybakov2019}. An important question is whether such homotopies can have physical relevance -- for example, whether they can represent dynamical trajectories or 
MEPs. The latter pass through SPs on the energy surface and 
provide an estimate of the activation energy
of transformations between magnetic configurations within HTST and Kramers/Langer theories. It is of particular interest to determine whether homotopy MEPs always exist and whether their energy barriers are smaller than those involving discontinuous transformations. This question is nontrivial since magnetization discontinuities have finite energy~\cite{heil2019,derras2019}, and the discrete nature of the system introduces further complexity. The SP search framework developed here is agnostic to whether an SP corresponds to a continuous or discontinuous transformation and is therefore well suited for addressing this problem.

\section{Method}
\label{sec:method}
The present work introduces a saddle-point search framework (SPSF) for magnetic systems. The framework is designed for automated and repeated application to different metastable states, and can be integrated into global optimization schemes and long-timescale dynamics simulations.

\begin{figure}
    \centering
    \includegraphics[width=1\linewidth]{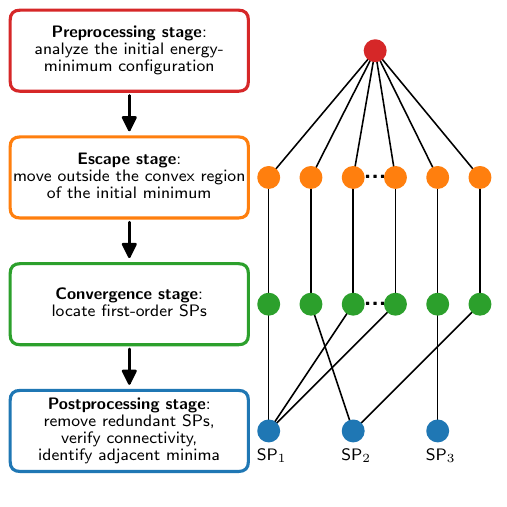}
    \caption{
    Overview of the saddle-point search framework (SPSF). The tree structure on the right illustrates that multiple SP searches can be initiated in parallel from a local energy minimum.
    }
    \label{fig:method_overview}
\end{figure}

The SPSF consists of four stages, illustrated in Fig.~\ref{fig:method_overview}: \textbf{preprocessing}, \textbf{escape}, \textbf{convergence}, and \textbf{postprocessing}. SP searches are initialized at local energy minima which may correspond to one or several localized magnetic textures. 
The geometric and symmetry properties of the textures are determined in the preprocessing stage (Sec.~\ref{ssec:method_preprocess}). This information is then used in the escape stage (Sec.~\ref{ssec:method_escape}) to generate various localized deformations that move the system toward the boundary of the convex region associated with the initial minimum. The resulting configurations serve as independent initial states for parallel calculations in the convergence stage (Sec.~\ref{ssec:method_find}), where the position of first-order SPs is obtained. Finally, in the postprocessing stage (Sec.~\ref{ssec:method_postprocess}), non-equivalent SPs are identified and their connectivity to the initial minimum is determined.

\subsection{Preprocessing stage}
\label{ssec:method_preprocess}
Figure~\ref{fig:preprocess_example} summarizes the steps involved in the preprocessing stage. 
%
First, the textures corresponding to an energy minimum are separated from one another and extracted from the FM background (see Sec.~\ref{sssec:object_recog}). 
Next, 
the symmetry elements and fundamental domains of the textures are determined using a dedicated symmetry-analysis algorithm (see Sec.~\ref{sssec:symmetry} and Sec.~\ref{sssec:fdomain}). Finally, the textures are passed to the escape stage. There, the fundamental domains are used to define various subsystems whose excitations guide the escape from the convex region of the energy minimum. 
The individual steps of the preprocessing stage are described in the following.

\subsubsection{Recognition of isolated magnetic textures}
\label{sssec:object_recog}
\begin{figure}
    \centering
    \includegraphics[width=1.0\linewidth]{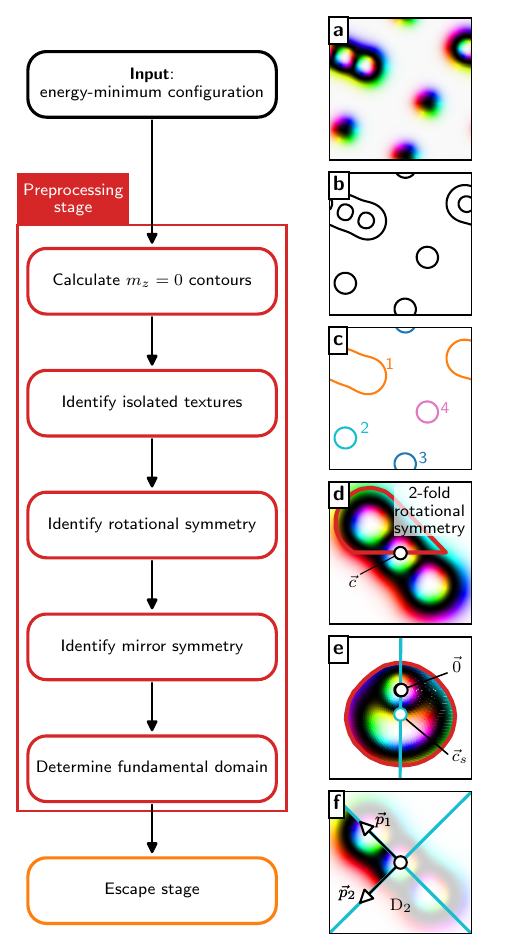}
    \caption{
    Flowchart of the preprocessing stage. The individual steps are illustrated in panels \textbf{a}-\textbf{f} for a magnetic state at $h=0.65$, $L_D=64a$, and $\beta=30^\circ$. \textbf{a}: Magnetic configuration corresponding to a local energy minimum. \textbf{b}: $m_z=0$ contours. \textbf{c}: Outer contours, each corresponding to an isolated magnetic texture. \textbf{d}: Selected texture with two-fold rotational symmetry; the red line indicates the boundary of the irreducible sector, and $\vec{c}$ denotes the center of mass of the texture.  \textbf{e}: Auxiliary texture obtained from the irreducible sector; its mirror axis (cyan line) is determined from the center of mass $\vec c_s$. 
    \textbf{f}: Fundamental domain of the original texture, bounded by two adjacent mirror axes $\vec{p}_1$ and $\vec{p}_2$ (cyan lines). The texture is identified as having $\mathrm{D}_2$ symmetry.}
    \label{fig:preprocess_example}
\end{figure}

A metastable configuration may contain several localized textures; an example involving three skyrmions and one texture with three inner contours is shown in Fig.~\ref{fig:preprocess_example}a. 
The textures are isolated from the FM background using the distribution of the out-of-plane magnetization component $m_z$.
In particular, the $m_z=0$ contours are computed, and the lattice sites located within one nearest-neighbor distance from the contours 
%
%
are identified and grouped into clusters using the density-based clustering algorithm~\cite{ester1996dbscan,schubert2017dbscan} with a metric accounting for periodic boundary conditions (see Fig.~\ref{fig:preprocess_example}b). 
Each cluster is associated with a single closed contour, for which the convex hull is computed~\cite{gamby2018,jarvis1973convexhull}. 
Possible inner contours are detected by testing whether their corresponding convex hulls are enclosed by another contour. 
Finally, isolated magnetic textures are determined from the outermost disjoint contours, with the magnetic moments enclosed by each such contour defining a single texture (see Fig.~\ref{fig:preprocess_example}c).

The center of mass of a given isolated texture can be computed as~\cite{Ivanov2005,Sheka2006,Moutafis2009,Makhfudz2012}:
\begin{align}
\vec{c}=\frac{\sum\limits_{i\in\mathcal{I}}\vec{r}_{i}(m_{i}^z-1)}{\sum\limits_{i\in\mathcal{I}}(m_{i}^z-1)}~,
\label{eq:fixpoint1}
\end{align}
where $\mathcal{I}$ is the set of sites associated with the texture, and $\vec{r}_{i}$ is the position of site $i$. 

\subsubsection{Recognition of symmetry}
\label{sssec:symmetry}
The construction of fundamental domains for isolated magnetic textures relies on identifying the symmetry elements of the textures. These are described by elements $[\mathcal{M}||\mathcal{N}]$ of spin point groups~\cite{Litvin1977,liu2022,xiao2024,Schiff2025}, where $\mathcal{N}$ acts in real space and $\mathcal{M}$ acts in the space of magnetic moments:
\begin{align}
    [\mathcal{M}||\mathcal{N}]\vec{m}(\vec{r}_i)=\mathcal{M}\vec{m}(\mathcal{N}^{-1}(\vec{r}_i-\vec{c})+\vec{c})~,
\end{align}
where $i\in\mathcal{I}$ labels the lattice sites associated with the texture. 
An element $[\mathcal{M}||\mathcal{N}]$ is a symmetry of the texture if $[\mathcal{M}||\mathcal{N}]\vec{m}(\vec{r}_i)=\vec{m}(\vec{r}_i)$ for all $i\in\mathcal{I}$. 
In the present application, the relevant spin point group elements are rotations and reflections. The rotation operations act as~\cite{kuchkin2021}:
\begin{align}
[R_{2\pi/n}||R_{2\pi/n}]\vec{m}(\vec{r}_i)=R_{2\pi/n}\vec{m}\left(R_{-2\pi/n}(\vec{r}_i-\vec{c})+\vec{c}\right)~.\label{eq:rotation}
\end{align} 
Textures invariant under such $n$-fold rotations possess the spin point group symmetry $\mathrm{C}_n$. 
The reflections are given by~\cite{barton2020}:
\begin{align}
    [R_{\pi-2\beta}P||P]\vec{m}(\vec{r}_i)=R_{\pi-2\beta}P\vec{m}(P(\vec{r}_i-\vec{c})+\vec{c})~,\label{eq:generator_reflection}
\end{align}
where $P=2\vec{p}\vec{p}^T-I$ is the reflection matrix, with $\vec{p}$ denoting the unit vector along the mirror axis in the system plane and $I$ the $3\times 3$ identity matrix. The additional rotation by $\pi-2\beta$ in spin space accounts for the helicity of the energy model. The symmetry of textures invariant under $n$-fold rotations combined with $n$ reflections is described by the spin point group $\mathrm{D}_n$. Figure~\ref{fig:introduction_textures} shows examples of magnetic textures corresponding to different spin point groups. Axially symmetric textures such as skyrmions (Fig.~\ref{fig:introduction_textures}b) correspond to the infinite group $\mathrm{D}_\infty$. 

When testing whether Eq.~\eqref{eq:rotation} or Eq.~\eqref{eq:generator_reflection} represents a symmetry, it is important to realize that the transformed magnetic moment positions generally do not coincide with the original lattice sites. Therefore, cubic interpolation is used in practice to evaluate the transformed magnetic moments at the original positions. The similarity between the transformed and original textures is quantified using the geodesic distance~\cite{bessarab2015}. A symmetry operation is considered present if the geodesic distance per magnetic moment is smaller than a certain threshold. 
Two- to eleven-fold rotations are tested as candidate symmetry operations. If all satisfy the symmetry criterion, the texture is classified as possessing continuous rotational symmetry. 

Once the rotational symmetry of the texture has been assessed, possible mirror symmetries are tested. For a texture with continuous rotational symmetry, any in-plane line passing through its center of mass can be used as a candidate mirror axis. The texture is classified as $\mathrm{D}_\infty$ if the corresponding mirror symmetry is present and as $\mathrm{C}_\infty$ otherwise.

For a texture with finite $n$-fold rotational symmetry, mirror symmetries are assessed using an irreducible sector of the texture. The sector is defined as the region between two radial directions from the texture center of mass $\vec c$, separated by the angle $2\pi/n$. Defining $\varphi_i$ as the polar angle of $\vec r_i-\vec c$ measured from the $x$-axis, the sector is chosen to contain all sites satisfying $0 \le \varphi_i \le 2\pi/n$. The irreducible sector of a texture with 
$2$-fold rotational symmetry is shown in Fig.~\ref{fig:preprocess_example}d.

The selected sector is then transformed into an auxiliary texture by angular rescaling so that it spans the full $2\pi$ range (see Fig.~\ref{fig:preprocess_example}e for an example). Specifically, the magnetic moments and lattice positions are transformed according to
\begin{align}
        \vec{m}_i^s=&R_{(n-1)\varphi_i}\vec{m}_i\label{eq:trafo_m}\\
        \vec{r}_i^s=&R_{(n-1)\varphi_i}(\vec{r}_i-\vec{c})
        \label{eq:trafo_r}
\end{align}
for all sites of the sector.


The auxiliary texture can have at most one mirror axis. The axis must pass through the origin, which corresponds to the center of mass $\vec c$ of the original texture [see Eq.~\eqref{eq:trafo_r}]. Its orientation is therefore fixed by one additional point invariant under reflection. This point is chosen as the center of mass of the auxiliary texture, denoted by $\vec c_s$. The direction of the candidate mirror axis is then given by $\vec{p}_s=\vec{c}_s$.

If the auxiliary texture does not exhibit mirror symmetry, the original texture has no mirror axes and is classified as having $\mathrm{C}_n$ symmetry. 
If mirror symmetry is present, the original texture is classified as having $\mathrm{D}_n$ symmetry and has $n$ mirror axes, $\vec{p_1},\dots,\vec{p}_n$, whose orientation is given by 
    \begin{align}
        \kappa_l=\frac{(l-1)\pi}{n}+\frac{\kappa_s}{n}, \quad l=1,\dots,n,
    \end{align}
where $\kappa_s$ and $\kappa_l$ are the angles of $\vec{p}_s$ and $\vec{p}_l$, respectively, measured from the $x$-axis. The corresponding mirror axes are illustrated for a texture classified as having $\mathrm{D}_2$ symmetry in Fig.~\ref{fig:preprocess_example}f.
\subsubsection{Determination of fundamental domains}
\label{sssec:fdomain}
The fundamental domain of a texture is determined from the identified symmetry elements (see Fig.~\ref{fig:preprocess_example}f for a texture with $\mathrm{D}_2$ symmetry). For textures with $\mathrm{C}_\infty$ and $\mathrm{D}_\infty$ symmetry, the fundamental domain is chosen as the radial line segment connecting the texture center of mass to the outer contour along the $y$-direction. 

For textures with finite $n$, the fundamental domain is chosen as a sector of the texture bounded by the outer contour and two radial line segments extending from the texture center $\vec{c}$. For a texture with $\mathrm{C}_n$ symmetry and no mirror axes, the angle between the line segments is $2\pi/n$. For a texture with $\mathrm{D}_n$ symmetry and $n>1$, the line segments are oriented along adjacent mirror axes.  
For $\mathrm{D}_1$ symmetry, the fundamental domain is chosen as one of the two halves of the texture separated by the mirror axis. 
Fundamental domains for several textures are illustrated in Figs.~\ref{fig:introduction_textures}c-f.

\subsection{Escape stage}
\label{ssec:method_escape}
The GMMF method~\cite{muller2018,schrautzer2025} and related approaches~\cite{bocquet2023} used in the convergence stage of the SP search framework should not be initialized from within the convex region of a local minimum.
Suitable starting points for the convergence stage are generated during the escape stage, and this task is nontrivial.
The objective is to sample a diverse set of points outside the convex region, where at least one eigenvalue of the Hessian becomes negative.
Ideally, these points should lie in different regions containing ascending valleys so as to maximize the likelihood that the subsequent convergence stage, initialized at these points, will reach distinct SPs.
Therefore, the escape stage is a crucial part of the SPSF~\cite{plasencia2014,gunde2024}, and its strategy must be chosen carefully so as to maximize the completeness of SP sampling.

\begin{figure*}
    \centering
    \includegraphics[width=1\linewidth]{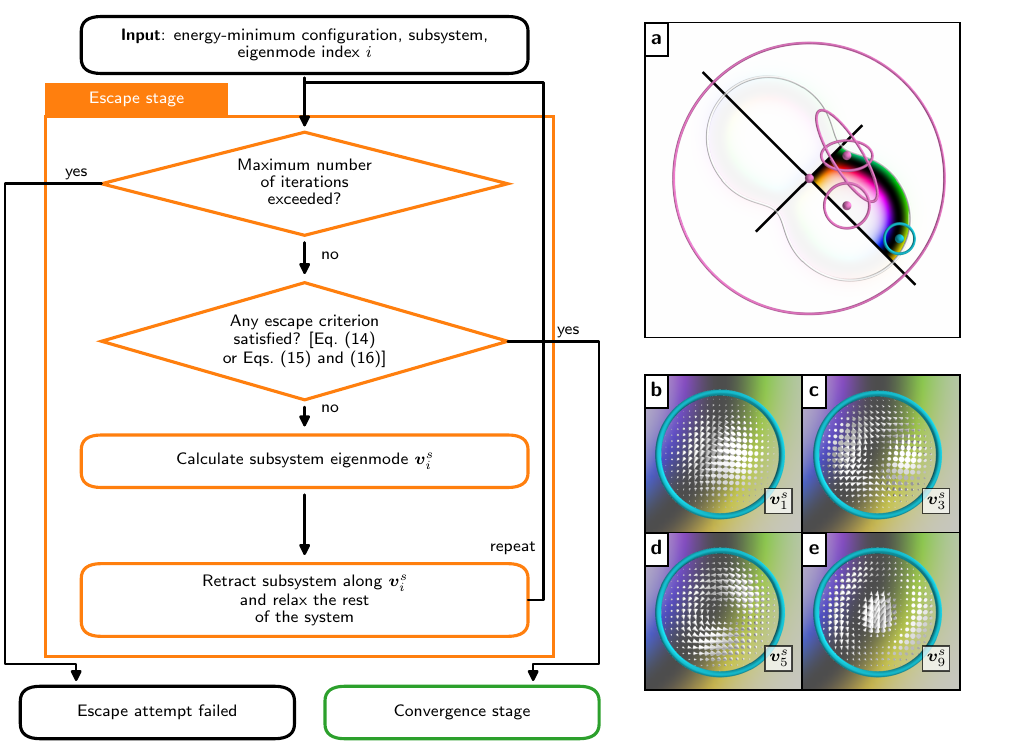}
    \caption{Flowchart of the escape stage. \textbf{a}: Sampling of elliptical subsystems centered within the fundamental domain of a magnetic texture. For the subsystem highlighted in cyan, some of the low-energy eigenmodes used to guide the escape attempts are shown in \textbf{b}--\textbf{e}. Each sampled subsystem and eigenmode defines an independent escape attempt, and both signs of the corresponding eigenvector are considered.
%
%
%
}
    \label{fig:flowchart_escape}
\end{figure*}

An escape strategy proposed in this study is based on low-energy excitations of subsystems of a given texture.
Each subsystem is defined as an ellipse-shaped region centered within the fundamental domain of the texture. 
A family of about 50 subsystems is generated by sampling the position of the ellipse center, the length of the semi-axes, and the orientation of the major axis (see Fig.~\ref{fig:flowchart_escape}a).
This sampling ensures that different physically nonequivalent parts of the texture are probed.
Note that ellipses may extend beyond the fundamental domain.
In particular, at least one subsystem is defined to encompass the entire texture, ensuring that deformations affecting the full texture are also captured.

Starting from an energy-minimum state, escape trajectories are generated by concerted rotations of the magnetic moments, guided by the low-energy eigenmodes of a subsystem, i.e., the eigenvectors of the corresponding partial Hessian.
In practice, each escape trajectory is constructed iteratively.
At each iteration, the magnetic moments in the subsystem are displaced by a small distance using a retraction~\cite{edelman1998,schrautzer2025}, after which the rest of the system is relaxed using the \textit{limited-memory Broyden–Fletcher–Goldfarb–Shanno} (L-BFGS) method~\cite{ivanov2021}.
This energy minimization while keeping the subsystem configuration fixed ensures smooth magnetization across the subsystem boundary and prevents formation of artificial boundary effects.
After this, the subsystem eigenmode is updated, providing input for the next iteration.
The procedure of generating an escape trajectory is summarized in the flowchart in Fig.~\ref{fig:flowchart_escape}.

An escape attempt is considered successful if the iterative procedure brings the system to a point outside the convex region of the initial minimum. 
In particular, an attempt is accepted when the following condition is satisfied:
\begin{equation}
    \lambda_1\leq\lambda_\text{esc},\label{eq:escape_crit_single}
\end{equation}
where $\lambda_1$ is the lowest eigenvalue of the Hessian $H$ and the threshold parameter $\lambda_\text{esc}<0$ controls how far the system must move away from the convex region. 
This escape threshold is particularly important in high-dimensional systems, where configurations can exhibit soft modes, along which an excitation corresponds to overcoming a very small, but different from zero, energy barrier.
During the escape process, such a mode can acquire a small negative eigenvalue indicating the vicinity to an SP associated to e.g. a skyrmion translation~\cite{malottki2019,bessarab2018}.
An escape threshold efficiently filters out these uninteresting SPs.

If only the condition in Eq.~\eqref{eq:escape_crit_single} is enforced, there is a risk of reentering the convex region during the convergence stage, as shown in Appendix~\ref{sec:app_escape}.
Therefore, the condition of being outside the convex region ($\lambda_1<0$)
can be supplemented by 
requiring sufficient alignment between the energy gradient and the minimum mode, which is a signature of ascending valleys on the energy surface~\cite{hoffman1986,jorgensen1988,schlegel1992,bofill2012}. This yields an alternative escape condition:
\begin{align}
    \lambda_1&<0,\label{eq:escape_crit_fullsys}\\
    \frac{|\bm{v}_1\cdot\bm{g}|}{|\bm{g}|}&\geq w_\text{esc},\label{eq:valley_entrance}
\end{align}
where 
$\bm{v}_1$ is the eigenvector corresponding to the lowest eigenvalue of the Hessian, also referred to as the minimum mode, and $\bm{g}=(\vec{g}_1,\dots,\vec{g}_N)$ is the gradient of the energy with respect to the orientation of the magnetic moments:
\begin{align}
    \vec{g}_i = \frac{\partial E}{\partial{\vec{m}_i}}-\left(\frac{\partial E}{\partial{\vec{m}_i}}\cdot\vec{m}_i\right)\vec{m}_i~.
    \label{eq:gradient}
\end{align}
Note that the Hessian $H\in\mathbb{R}^{2N\times 2N}$ and the gradient adhere to the curvature of the $2N$-dimensional configuration space of magnetic systems~\cite{muller2018,varentcova2020,bocquet2023,schrautzer2025}.

The choice of the escape thresholds affects the starting configurations for the convergence stage and, consequently, the likelihood of reaching first-order SPs rather than ending up in convex regions. How the outcome of the convergence stage depends on the threshold values is analyzed in Appendix~\ref{sec:app_escape}. For the system considered in this study, we use $\lambda_\text{esc}=-0.2\mathcal{J}$ and $w_\text{esc}=0.25$.


In practice, it is sufficient to apply Eq.~\eqref{eq:escape_crit_single} or Eqs.~\eqref{eq:escape_crit_fullsys},\eqref{eq:valley_entrance} 
to a subsystem. 
Consider the partial Hessian $H^s$ formed by selecting the rows and columns of $H$ corresponding to the magnetic moments of the subsystem. Since $H^s$ is a compression of $H$, 
Cauchy’s interlacing eigenvalue theorem~\cite{hwang2004cauchy} applies and gives:
\begin{align}
    \lambda_{1}\leq\lambda_1^s~,
    \label{eq:cauchy_theorem}
\end{align}
where $\lambda_1^s$ is the lowest eigenvalue of the subsystem Hessian. 
Hence, checking $\lambda_{1}^s\leq\lambda_\text{esc}$ and $\lambda_1^s<0$ is a conservative and computationally inexpensive way of testing the criteria in  Eqs.~\eqref{eq:escape_crit_single},\eqref{eq:escape_crit_fullsys}.
After a few escape-stage iterations, 
the minimum mode of the full system becomes strongly localized within the chosen subsystem. 
Therefore, it is sufficient to evaluate the escape criterion in Eq.~\eqref{eq:valley_entrance} by calculating $|\bm{v}_1^s\cdot \bm{g}^s|/|\bm{g}^s|\geq w_\text{esc}$, where $\bm{g}^s$ and $\bm{v}_1^s$ denote the energy gradient and the minimum mode of the subsystem, respectively.

A set of suitable starting points for the convergence stage is obtained by sampling the parameters of the subsystems and the low-energy eigenmodes of each subsystem. In practice, it is sufficient to sample deformations described by only the lowest five to ten eigenmodes, taking into account both signs of the corresponding eigenvectors. 
Fig.~\ref{fig:flowchart_escape}a shows an example of sampling over different subsystems centered within the fundamental domain of the double-bag state. Figs.~\ref{fig:flowchart_escape}b-e visualize four subsystem eigenmodes that can be used in the escape stage.

The proposed escape-stage strategy provides control over the localization and symmetry of the generated deformations. In particular, it enables targeting SPs associated with transformations that locally break the symmetry of the initial magnetic texture. This is illustrated by the SP corresponding to the skyrmion duplication~\cite{muller2018,schrautzer2025}, where a configuration with $\mathrm{D}_\infty$ symmetry transforms via a figure-eight-shaped SP configuration exhibiting $\mathrm{D}_2$ symmetry. In approaches based on eigenmodes of the full system, introducing a tendency toward such an SP may require successive switching between multiple modes~\cite{schrautzer2025}. In contrast, sampling spatially confined subsystem excitations allows direct access to such symmetry-breaking deformations.

In addition, the proposed strategy reduces the computational cost of the escape stage compared to approaches based on the full system. Recomputing the Hessian spectrum for the full system is expensive. The subsystems contain an order of magnitude fewer magnetic moments, allowing many escape attempts to be performed efficiently and in parallel.

\subsection{Convergence stage}
\label{ssec:method_find}
The escape stage yields a set of magnetic configurations outside the convex region of the initial energy minimum. These configurations are then used to initialize the convergence stage, which aims to identify first-order SPs adjacent to that minimum.
A detailed description of the 
GMMF method~\cite{muller2018} used in the convergence stage as well as its efficient implementation can be found in Ref.~\cite{schrautzer2025}. Here, a brief summary is provided for completeness.
The GMMF method iteratively advances the magnetic structure so as to maximize the energy along a certain direction -- the inversion mode $\bm{q}$ -- while minimizing the energy along all other directions. 
In particular, the GMMF force guiding the rotation of the magnetic moments is obtained by reversing the component of the gradient $\bm{g}$ along the inversion mode:
\begin{align}
    \bm{f} = -\bm{g}+2(\bm{g}\cdot\bm{q})\bm{q}.
    \label{eq:GMMF_force_full}
\end{align}
In the vicinity of an SP, the inversion mode is always the minimum mode, i.e., $\bm{q}=\bm{v}_1$. However, near the boundary of the convex region and/or in case of a mode-crossing event, choosing the inversion mode to be the Hessian’s eigenvector corresponding to the second lowest
eigenvalue or even higher eigenvalue
can improve the performance of the GMMF method~\cite{schrautzer2025}.

The convergence on an SP is achieved using some numerical optimization technique, preferably respecting the curvature of the magnetic configuration space~\cite{edelman1998,schrautzer2025}, equipped with the GMMF force. We find that an 
L-BFGS algorithm adapted for magnetic systems~\cite{ivanov2021} yields optimal performance.
A GMMF calculation is considered converged on a first order SP if the following conditions are met simultaneously:
\begin{equation}
    \max_{i\in\{1,\dots,N\}}|\vec{f}_{i}|\leq f_\text{conv},~~\lambda_1<0,~~\text{and}~~\lambda_2\geq 0~,
    \label{eq:conv_crit_GMMF}
\end{equation}
where $f_\text{conv}=10^{-10}$~eV/rad is used in this study.

During a calculation, the inversion mode $\bm{q}$ and the gradient $\bm{g}$ may become nearly orthogonal, causing the GMMF force to reduce to $-\bm{g}$.
In this case, the GMMF optimization essentially becomes an energy minimization, which may pull the system into a convex region and prevent convergence on an SP. Such attempts are regarded as failed.

The computational bottleneck of the GMMF algorithm -- the repeated calculation of the inversion mode -- can be addressed efficiently via direct minimization of the generalized Rayleigh Quotient using the L-BFGS solver~\cite{liu1989}, adapted
to account for the curvature of the Grassmann manifold on which the objective function is defined, while avoiding explicit Hessian evaluations~\cite{schrautzer2025}. 
Efficient evaluation of the inversion mode is significant especially for large systems and/or systems with long-range interactions. 
For details of the implementation, see Ref.~\cite{schrautzer2025}. 

Performance of the convergence stage can be further improved by applying the GMMF method to  subsystems defined in the escape stage. In such subsystem-based GMMF (S-GMMF) approach, only magnetic configuration of the subsystem is modified iteratively according to the GMMF algorithm, while the rest of the system is relaxed after each iteration. Computing the inversion mode for a subsystem is computationally less demanding than for the full system. On the other hand, the subsystem-constrained SP found using the S-GMMF method is often quite close to the true SP of the full system, allowing the subsequent GMMF calculation applied to the full system to converge quickly. This formulation significantly reduces the overall computational cost of the convergence stage compared with applying GMMF to the full system immediately after the escape stage.
Another benefit of the S-GMMF approach is that, by constraining the SP search to the same subsystems used in the escape stage, the method improves the correspondence between escape-stage excitations and the identified SPs, making SP sampling more systematic.
Every S-GMMF attempt is followed by a full-system GMMF calculation, initialized with the final configuration of the S-GMMF calculation, even if the S-GMMF does not find a subsystem-constrained SP.
S-GMMF may fail to converge if the chosen subsystem is too small: In such cases, following the S-GMMF force typically leads into a convex region of the subsystem. However, according to Eq.~\eqref{eq:cauchy_theorem}, this does not imply that the corresponding configuration lies in the convex region of the full system. As a result, the subsequent full-system GMMF calculation often converges on an SP.

\begin{figure*}
    \centering
    \includegraphics[width=1.0\linewidth]{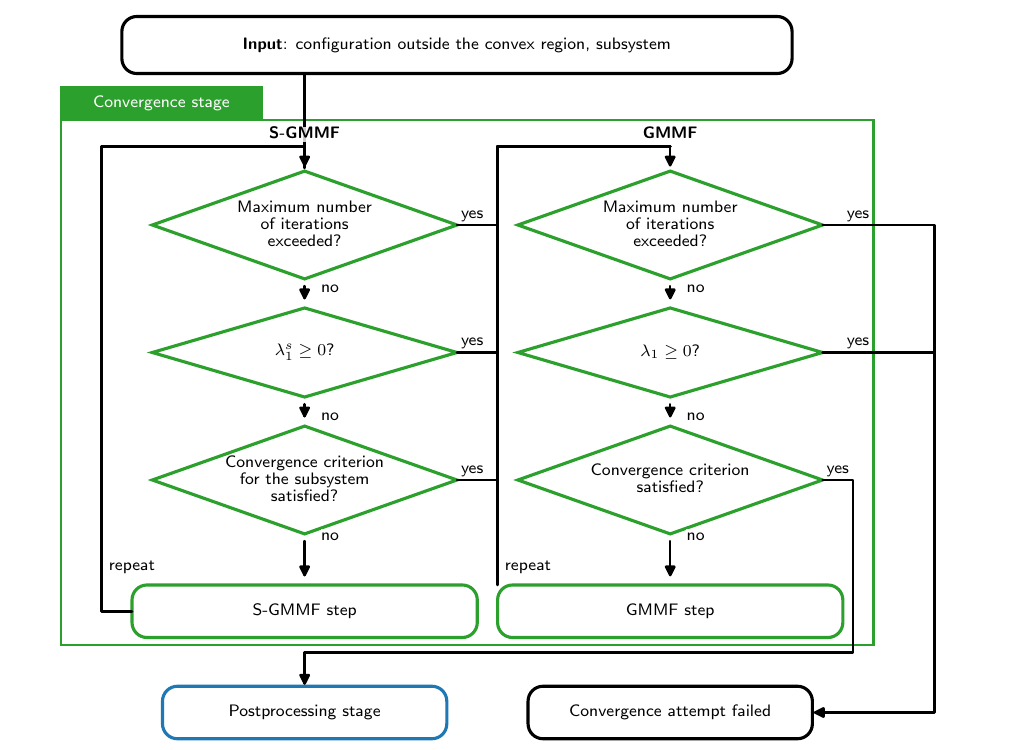}
    \caption{Flowchart of the convergence stage.
    }
    \label{fig:method_flowchart_find}
\end{figure*}

A flowchart summarizing the convergence stage is shown in Fig.~\ref{fig:method_flowchart_find}.

\subsection{Postprocessing stage}
\label{ssec:method_postprocess}
Different SP searches may converge to SPs that represent essentially the same magnetic configuration, differing only by a translation and/or a global rotation of the texture. Consequently, the list of SPs obtained during the  convergence stage can contain significant redundancy. The purpose of the postprocessing stage is to eliminate this redundancy, yielding a unique set of saddle points, and to embed them into a state network by identifying the adjacent energy minima.

\subsubsection{Filtering of redundant saddle points}
\label{sssec:method_clustering}
Cluster algorithms used in unsupervised learning are well suited to identify redundant SPs by grouping the corresponding magnetic textures into clusters according to their similarity. 
Two configurations are considered similar when their feature vectors are close. 
We find it sufficient to characterize the magnetic textures using compact feature vectors whose components include total energy, individual interaction-resolved contributions to the total energy, the topological charge, and the two lowest eigenvalues of the Hessian.
Other choices of features are also possible.

To make the features comparable, each component of the feature vectors is normalized to the unit interval. A subsequent principal component analysis~\cite{pearson1901_pca1,jolliffe2002_pca2} reduces the feature vectors to two dimensions, yielding
$\{\bm{{\mathcal{F}}}_1,\dots,\bm{{\mathcal{F}}}_{N_\text{SP}}\}$ with $\bm{{\mathcal{F}}}_j\in\mathbb{R}^2$ and $N_\text{SP}$ being the number of SPs found in the convergence stage. These reduced feature vectors form the input to the clustering procedure, where two SPs are assigned to the same cluster if the norm of their feature-vector difference is small. SPs within the same cluster correspond to essentially equivalent magnetic configurations, whereas SPs in different clusters represent significantly different textures.

Because the number of distinct SPs is not known a priori, the number of clusters must be determined from the data. Given the modest number of SP configurations on the scale of machine learning problems, we repeatedly apply k-means clustering~\cite{Bock2007} for several candidate cluster counts and select the best partition
using the silhouette score~\cite{saputra2020kmeans}. 
This workflow can be implemented straightforwardly using standard machine-learning libraries~\cite{scikit-learn}. Only one representative from each cluster is kept, yielding a set of unique SP configurations. 

\subsubsection{Connectivity and adjacent minima}
\label{sssec:method_connectivity}
To construct the graph of metastable states, the energy minima adjacent to each SP must be determined. For this purpose, the system is displaced slightly from the SP in both directions along the unstable mode, and the resulting configurations are used as starting points for steepest-descent energy minimizations.
These minimizations determine whether the SP is connected to the initial minimum used in the SPSF search and identify the adjacent minimum, thereby extending the network of states.
If neither minimization returns to the initial minimum, the SP is still stored, as it corresponds to a transition between some other metastable states and therefore remains relevant for mapping the energy landscape. 

\section{Results}
\label{sec:results}
In the following, we apply the developed SPSF to the 2D chiral magnet system to investigate transitions between metastable magnetic textures formed by various combinations of closed loops, chiral kinks, and tails.
In Sec.~\ref{ssec:results_methodpresentation}, we analyze SPs surrounding the skyrmion state and identify the adjacent textures corresponding to local energy minima. 
SPSF calculations initiated from each minimum progressively map out the graph of magnetic textures, where the edges correspond to SPs and the vertices correspond to local energy minima.
In Sec.~\ref{sssec:results_graph} we analyze this graph and reveal a hierarchy of energy barriers associated with transformations of the fundamental elements of magnetic textures. 
In Sec.~\ref{sssec:scaling_isolated_stripe} we examine how these transformations change under variation of the lattice discretization.
Finally, Sec.~\ref{ssec:existenceMEP_homotopy} addresses whether an MEP always exists between two magnetic states belonging to the same homotopy class. 
Unless stated otherwise, the external magnetic field is set to $h=0.623$, for which all of the fundamental elements -- closed contours, chiral kinks, and tails -- can be present as components of meta-stable textures~\cite{kuchkin2022_phd}, and the period of the helical ground-state modulation is set to $L_D=40a$.
%
\subsection{Skyrmion collapse mechanisms and adjacent states}
\label{ssec:results_methodpresentation}
\begin{figure*}
    \centering
    \includegraphics[width=1.0\linewidth]{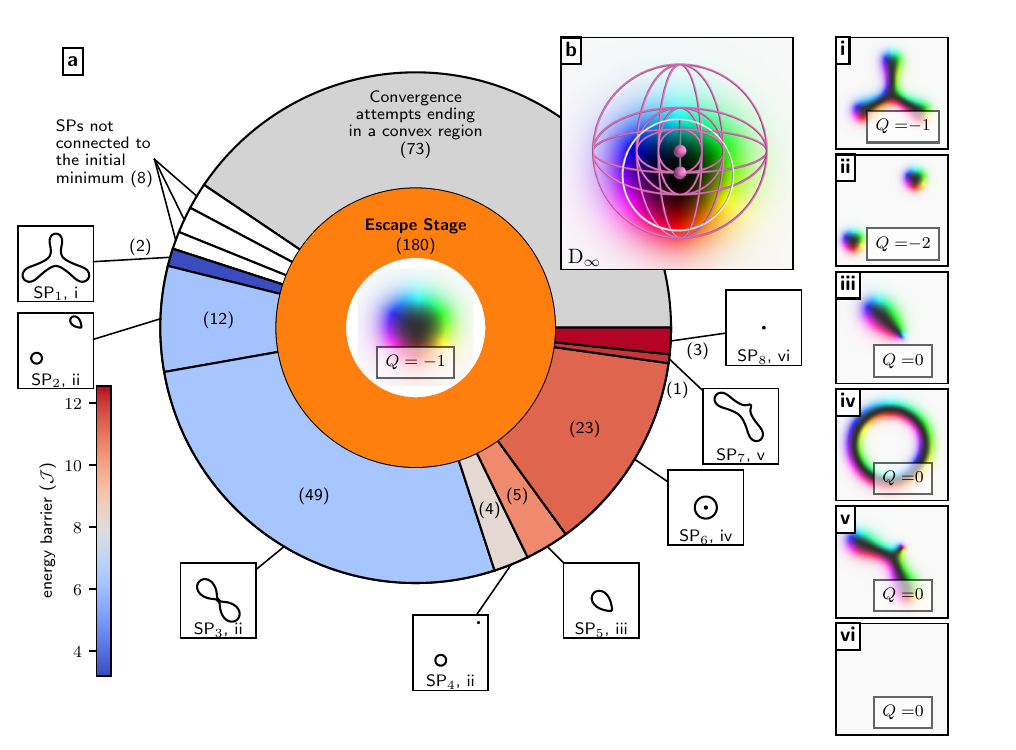}
    \caption{\textbf{a}: Diagram summarizing the results of the SPSF calculations applied to the skyrmion state ($\mathrm{D}_\infty$, $Q=-1$) at $h=0.623$ and $L_D=40a$. The inner orange ring represents the 180 escape attempts, all of which are successful. The outer sectors represent the outcomes of the subsequent convergence attempts; the sector size and the number in parentheses indicate how many attempts lead to the corresponding outcome. The gray sector corresponds to failed attempts ending in a convex region. The white sectors correspond to attempts yielding SPs not connected to the initial skyrmion-state minimum. Each colored sector represents attempts yielding a unique SP connected to the skyrmion-state minimum, with the color indicating the corresponding energy barrier, as shown by the color bar. The SP configurations are shown in the insets, and the configurations corresponding to the adjacent energy minima are shown in panels \textbf{i}-\textbf{vi}. \textbf{b}: Sampling of subsystem geometries used in the escape stage.
    }
    \label{fig:sk_sps}
\end{figure*}

Fig.~\ref{fig:sk_sps} provides an overview of the SPSF calculations applied to the most prominent 2D topological magnetic texture -- axially-symmetric skyrmion with topological charge $Q=-1$. 
Since the skyrmion exhibits $\mathrm{D}_\infty$ symmetry, its fundamental domain is a line segment from the skyrmion center to its perimeter contour (Fig~\ref{fig:sk_sps}b).
Two points on the skyrmion radius are chosen as centers of the elliptical subsystems used to generate excitations in the escape stage (Fig.~\ref{fig:sk_sps}b).
Each ellipse is oriented with one of the axes aligned along the skyrmion radius.
For each center position, a set of nine ellipses is defined using the following values of the semiaxes: $L_D/2,L_D/4,L_D/8$.
Escape attempts are performed for each subsystem using the eigenvectors of the subsystem Hessian corresponding to the five lowest eigenvalues. 
Considering both directions along each eigenvector, this gives a total of $180$ escape attempts, all of which are successful: in each case, either Eq.~\eqref{eq:escape_crit_single} or Eqs.~\eqref{eq:escape_crit_fullsys} and \eqref{eq:valley_entrance} are eventually fulfilled. 


Each escaped configuration is passed to the convergence stage. Among 180 convergence attempts, 73 failed, ending in convex regions, while 107 yielded SPs. 
Postprocessing grouped them into clusters corresponding to 11 unique SPs. 
Connectivity tests using steepest-descent calculations showed that three of the identified SPs are not connected with the initial minimum. This results in eight mechanisms (SP$_1$-SP$_{8}$) of the skyrmion collapse.
%
The unique SPs, their occurrence counts in repeated searches, and their energies relative to the skyrmion state minimum are shown in Fig.~\ref{fig:sk_sps}a. 

The lowest-barrier mechanism connects the skyrmion with the triangular-shaped texture carrying three tails (Fig.~\ref{fig:sk_sps}i)~\cite{kuchkin2023}. 
Here, the skyrmion elongates and transforms into a texture with two tails, while SP$_{1}$ corresponds to the nucleation of a third tail between the two ends of the elongated skyrmion. 
The topological charge does not change upon the formation of tails.

The three SPs -- SP$_{2}$, SP$_{3}$ and SP$_{4}$ -- all connect the skyrmion state to the two-skyrmion state ($Q=-2$) (Fig.~\ref{fig:sk_sps}ii) via three different mechanisms.
In particular, SP$_{3}$ corresponds to the skyrmion duplication mechanism reported in Ref.~\cite{muller2018}.
SP$_{4}$ corresponds to the formation of a second isolated skyrmion in the FM background via a Bloch point-like defect nucleation.
SP$_{2}$ describes the mechanism where the second skyrmion is formed from an isolated droplet-shaped configuration. 
In this case, the presence of the skyrmion destabilizes the droplet, so that the final state is two skyrmions rather than a skyrmion-droplet pair.
In contrast, a single skyrmion can transform into a topologically trivial, metastable chiral droplet (Fig.~\ref{fig:sk_sps}iii). The corresponding SP$_{5}$, referred to as the chimera SP~\cite{meyer2019,Muckel2021}, describes nucleation of the chiral kink at the skyrmion perimeter -- a process that changes the topological charge by one. 

SP$_{6}$ corresponds to the nucleation of an inner contour within the skyrmion via a Bloch point-like defect, thereby transforming the skyrmion into a skyrmion bag also referred to as skyrmionium. 
Since the outer contour and inner contour have opposite windings, the skyrmion bag carries zero topological charge (Fig.~\ref{fig:sk_sps}iv).
The skyrmion can also elongate and form a kink on its perimeter, as described by SP$_{7}$.
The adjacent topologically trivial state (Fig.~\ref{fig:sk_sps}v) can be described either as a chiral droplet with an attached tail or as an elongated skyrmion with a chiral kink.
Finally, the skyrmion can disappear via the usual radially-symmetric collapse mechanism described by SP$_{8}$ (Fig.~\ref{fig:sk_sps}vi).

\begin{figure*}
    \centering
    \includegraphics[width=1.0\linewidth]{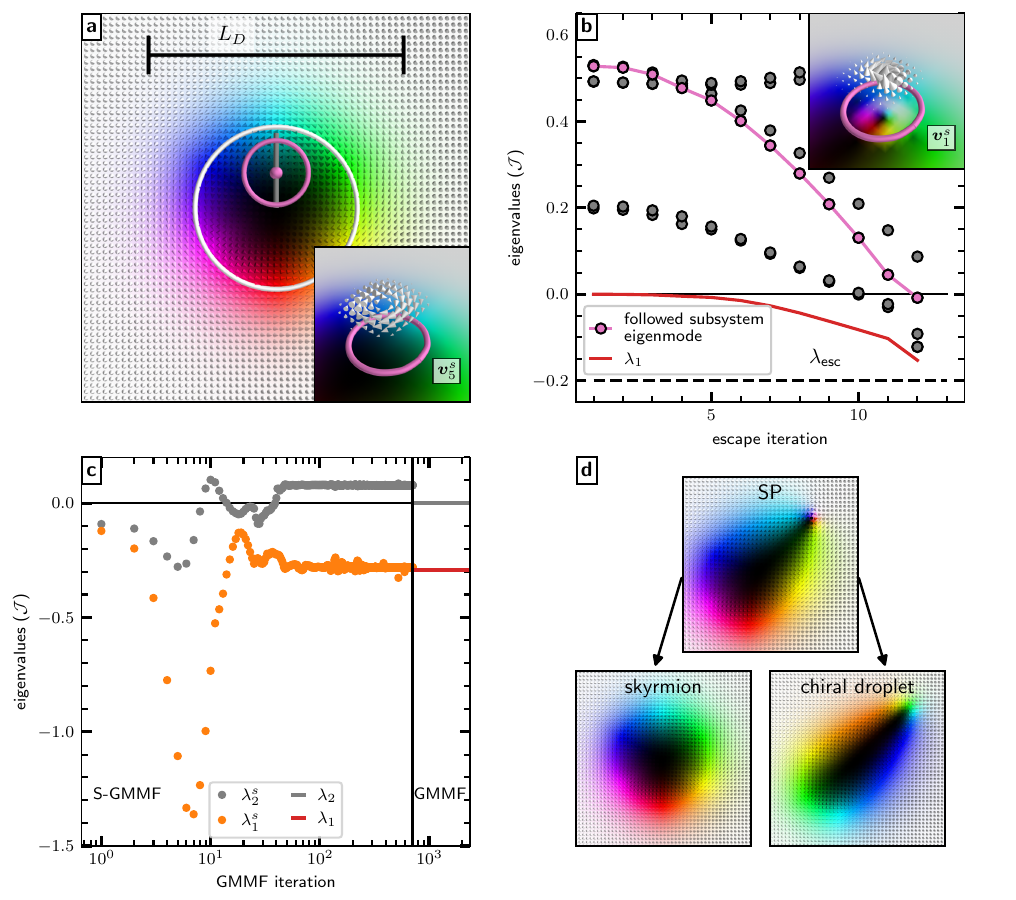}
    \caption{Example of a successful SP search initiated at the axially symmetric skyrmion state ($\mathrm{D}_\infty$, $Q=-1$) at $h=0.623$ and $L_D=40a$. \textbf{a}: Initial configuration with the circular subsystem used in the escape stage; the inset shows the subsystem eigenmode used to initiate the escape attempt. \textbf{b}: Evolution of the lowest eigenvalues of the subsystem and full-system Hessians during the escape stage; the threshold parameter $\lambda_\text{esc}$ is shown by the dashed line. The inset shows the subsystem eigenmode corresponding to the lowest eigenvalue of the subsystem Hessian at the end of the escape stage. Although the lowest eigenvalue does not fall below the threshold, the escape attempt is successful due to sufficient alignment between the energy gradient and minimum mode [see Eqs.~(\ref{eq:escape_crit_fullsys})-(\ref{eq:valley_entrance})]. \textbf{c}: Evolution of the lowest eigenvalues of the subsystem and full-system Hessians during the convergence stage. The vertical line marks the switch from S-GMMF to full-system GMMF. \textbf{d}: The SP configuration reached in the convergence stage and the two adjacent energy-minimum configurations identified by steepest-descent energy minimizations initiated from the SP in both directions along the unstable mode.
    }
    \label{fig:skyrmion_spstages}
\end{figure*}

Figure~\ref{fig:skyrmion_spstages} illustrates a successful SP search attempt using a circular subsystem with radius $L_D/8$ (Fig.~\ref{fig:skyrmion_spstages}a). During the escape stage, the deformation of the texture is guided by the subsystem eigenmode that, at the initial minimum, corresponds to the fifth-lowest eigenvalue of the subsystem Hessian (see the inset of Fig.~\ref{fig:skyrmion_spstages}a). Following the chosen eigenmode of the subsystem is accompanied by relaxation of the rest of the system, as described in Sec.~\ref{ssec:method_escape}. 
Note that mode crossings occurred several times during the escape stage, resulting in changes of the index of the eigenmode.
Fig.~\ref{fig:skyrmion_spstages}b presents several of the lowest eigenvalues $\lambda_{i}^s$ of the subsystem Hessian during the escape process.
After several iterations, the lowest eigenvalue becomes negative, while the gradient maintains a substantial overlap with the minimum mode, indicating a successful escape from the convex region and providing a suitable starting point for the subsequent convergence stage [see Eqs.~\eqref{eq:escape_crit_fullsys},\eqref{eq:valley_entrance}]. By the final iteration of the escape stage, the minimum mode of the subsystem and that of the full system describe nearly identical excitations. The escaped configuration with strongly canted magnetic moments and the vector field representation of the subsystem minimum mode are shown in the inset of Fig.~\ref{fig:skyrmion_spstages}b. 

The convergence stage, initialized from the escaped configuration, involves an S-GMMF calculation followed by a 
full-system GMMF refinement~
(Fig.~\ref{fig:skyrmion_spstages}c).
The application of the S-GMMF yields a subsystem-constrained SP corresponding to the formation of a chiral kink in the subsystem.
Subsequent full-system GMMF calculation converges to the chimera SP illustrated in Fig.~\ref{fig:skyrmion_spstages}d.
The lowest eigenvalues of the subsystem and full-system Hessians are nearly identical.
In contrast, the second-lowest eigenvalues differ qualitatively: in the subsystem, $\lambda_{2}^s>0$, while in the full system $\lambda_{2}\approx 0$, indicating a mode whose excitation costs almost no energy. This zero mode corresponds to a rotation of the texture shown in Fig.~\ref{fig:skyrmion_spstages}d around the chiral kink. Zero modes are absent in the subsystem due to the imposed constraints. 
%
%

Steepest descent calculations in the postprocessing stage confirm the connection of the identified SP to the initial skyrmion state and 
the 
chiral droplet state, which can be viewed as a skyrmion with a chiral kink (see Fig.~\ref{fig:skyrmion_spstages}d).

\begin{figure*}
    \centering
    \includegraphics[width=0.9\linewidth]{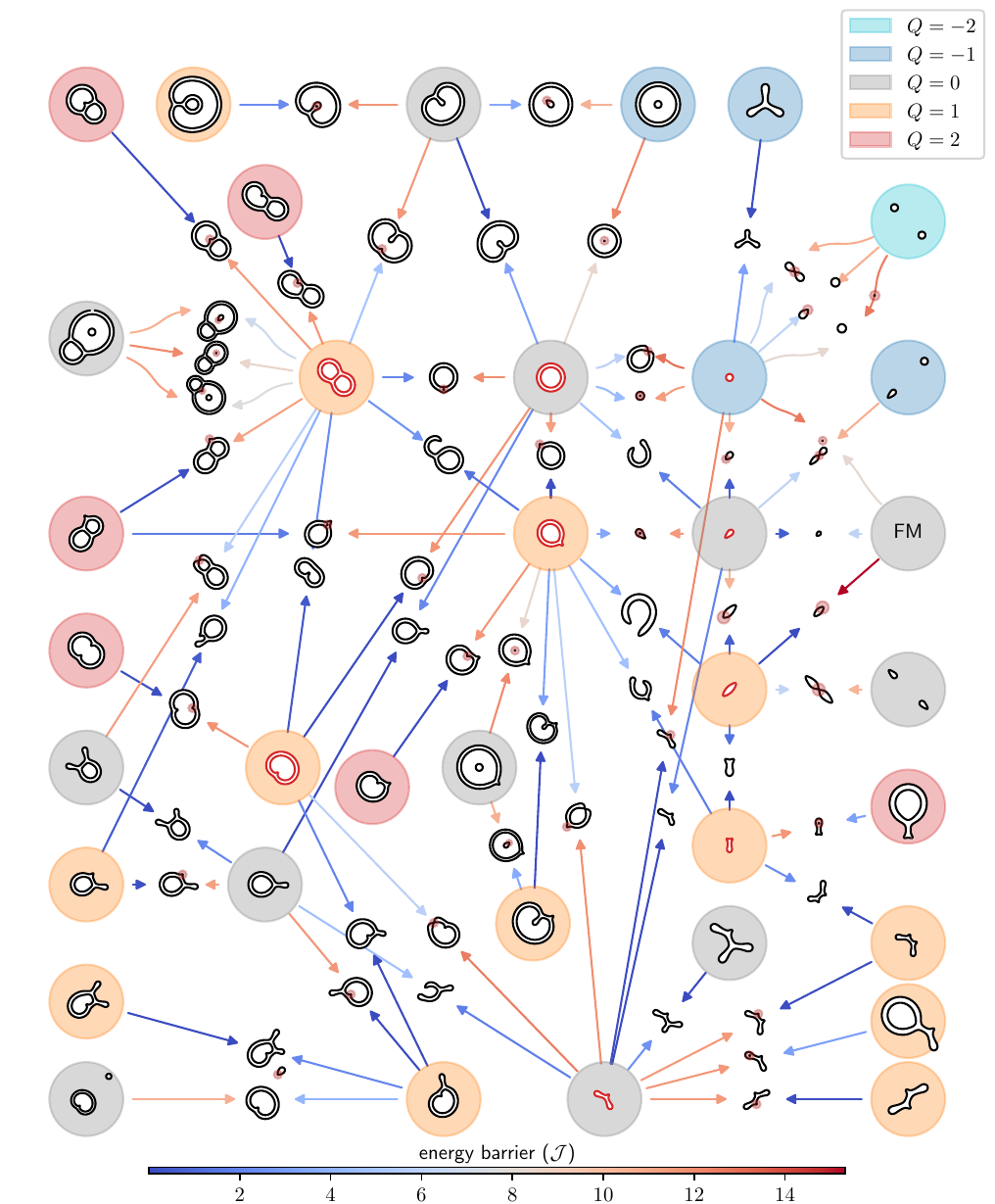}
    \caption{Network of metastable states revealed by repeated applications of the SPSF to a chiral magnet at $h=0.623$ and $L_D=40a$. The vertices of the graph represent metastable states, while the edges represent transitions between them, with the transition mechanisms defined by the corresponding SPs. Each edge is shown by two arrows pointing from the adjacent energy-minimum states toward the SP between them; the color of each arrow codes the corresponding energy barrier, as indicated by the color bar. Energy-minimum and SP configurations are represented by their $m_z=0$ contours. Red contours denote states from which SPSF searches were initiated, while black contours denote states obtained in subsequent applications of the framework. Semi-transparent red dots indicate the position of the maximum of $|\chi|$ in each SP configuration associated with a transition between states from different homotopy classes. The vertex colors code the topological charge $Q$ of the corresponding energy-minimum states, as indicated by the legend.
    %
   %
    }
    \label{fig:graph_LD40}
\end{figure*}


\subsection{Network of states in 2D chiral magnets}
\label{sssec:results_graph}
Recursive application of the SPSF to the metastable states enables the discovery of new metastable configurations and the connections between them, thereby allowing a systematic exploration of the energy surface of the magnetic system.
The resulting data can be represented as a directed graph in which energy minima form the vertices and the SPs defining transitions between the minima form the edges.

\begin{figure*}
    \centering
    \includegraphics[width=1.0\linewidth]{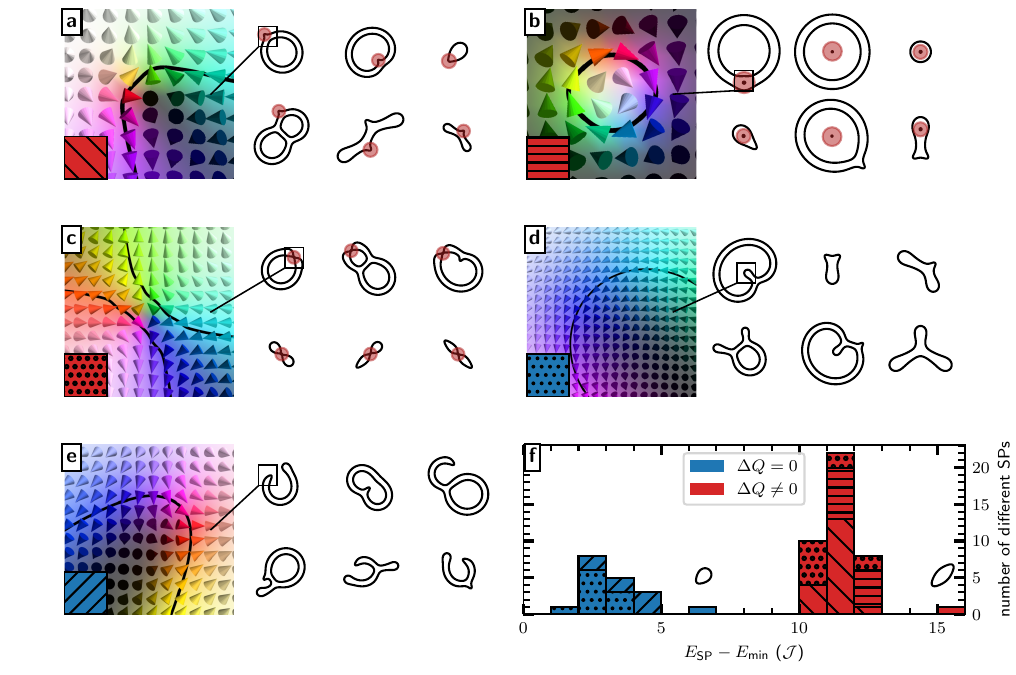}
    \caption{Mechanisms of texture transformations in a chiral magnet: nucleation of a chiral kink (\textbf{a}), formation of a closed contour through injection of a Bloch-point-like defect (\textbf{b}), contour merging (\textbf{c}), formation of a tail (\textbf{d}), and contour merging accompanied by nucleation of a chiral-kink (\textbf{e}). For each mechanism, six representative SP configurations are shown. Semi-transparent red dots indicate the position of the maximum of $|\chi|$ in each SP configuration associated with a transformation with $\Delta Q\ne 0$. \textbf{f}: Histogram of the SP energies $E_{\rm SP}-E_{\rm min}$, measured from the lower-energy of the two adjacent minima, for $h=0.623$ and $L_D=40a$. Blue and red bars correspond to transitions with $\Delta Q=0$ and $\Delta Q\neq 0$, respectively, while the hatching corresponds to the five transition mechanisms shown in \textbf{a-e}. SP configurations associated with the collapse of a chiral droplet and an antiskyrmion into the FM state are shown above the corresponding histogram bins.
%
 %
  %
    }
    \label{fig:histogram_LD40}
\end{figure*}

Figure~\ref{fig:graph_LD40} presents an example of such a graph, obtained after several recursive SPSF applications to the system with $L_D=40a$ and $h=0.623$ and labeled with the topological charge of each state and the energy barrier of each transition. For SP configurations connecting states from different homotopy classes, we also mark the position where the absolute value of the scalar spin chirality is largest. The scalar spin chirality is defined as~\cite{dosSantosDias2016,Grytsiuk2020}
\begin{align}
\chi_{ijk}=\vec{m}_i\cdot(\vec{m}_j\times\vec{m}_k)~,
\label{eq:spinchirality}
\end{align}
where $i$, $j$, and $k$ label the counterclockwise-ordered sites of an element in the triangulation of the square lattice. Large values of $|\chi|$ indicate strong noncoplanarity of neighboring magnetic moments, thereby indicating the locations of topological charge injection. 
Despite the complexity of the network of localized textures, most SPs characterizing transitions in this network can be grouped into five categories based on the transformations of the fundamental building blocks of the textures:
\begin{itemize}
    \item Nucleation of a chiral kink
    \item Formation of a closed contour via injection of a Bloch point–like defect
    \item Merging of two contours
    \item Formation of a tail
    \item Merging of two contours accompanied by the nucleation of a chiral kink
\end{itemize}
Examples of SP configurations from each category are shown in Fig.~\ref{fig:histogram_LD40}a-e.

The nucleation of a chiral kink (see Fig.~\ref{fig:histogram_LD40}a) locally changes the winding of the magnetization along a contour and thereby leads to small region where the sense of rotation of the magnetic moments is opposite to the systems's intrinsic chirality.
Since anisotropy is not included in this work, only positive chiral kinks, which contribute $+1$ to the winding number of the hosting contour, appear as constituents of metastable textures. 
Consequently, this SP category corresponds to transformations in which the total topological charge changes by $\Delta Q=+1$.

In contrast to the chiral kink nucleation, formation of an additional closed contour via injection of a Bloch-point–like defect (see Fig.~\ref{fig:histogram_LD40}b) can either increase or decrease the topological charge, $\Delta Q=\pm1$, 
depending on the winding of the newly formed contour. For example, formation of the contour in the FM vacuum decreases $Q$ by one, while nucleation of the contour inside the skyrmion perimeter increases $Q$ by one. 

SPs corresponding to the contour-merging capture the mechanism in which two closed contours -- nested or disjoint -- combine and form one closed contour (see Fig.~\ref{fig:histogram_LD40}c).
Similar to the Bloch point-like defect injection, such a transformation can either increase or decrease the total topological charge, $\Delta Q=\pm 1$, depending on the hierarchy of the merging contours as well as on their intrinsic winding, i.e. winding of the magnetization along the bare (kink-free) contours. In particular, merging two disjoint contours with positive (negative) intrinsic winding decreases (increases) $Q$ by one, while merging a contour with positive (negative) intrinsic winding with an inner contour increases (decreases) $Q$ by one. 




In contrast to chiral kink nucleation, Bloch point-like defect injection, and contour merging, the last two mechanisms correspond to homotopies, i.e. continuous transformations of the magnetization that preserve the topological charge of the system, $\Delta Q = 0$.
Indeed, formation of a tail (see Fig.~\ref{fig:histogram_LD40}d) only deforms a contour without changing sense of rotation of magnetization, thereby leaving the topological charge unchanged.

The SPs from the fifth category describe the second homotopy, where two closed contours merge simultaneously with the nucleation of a chiral kink (see Fig.~\ref{fig:histogram_LD40}e). This compensation ensures that the total topological charge remains unchanged. Such transformations occur only for disjoint contours with positive intrinsic winding or for nested contours in which the outer contour has negative winding. 
Examples of this mechanism include the transformation of a skyrmion bag into a chiral droplet, where the merging of two nested contours contributes $-1$ to the topological charge change, which is compensated by the nucleation of a chiral kink, and the transformation of a skyrmion bag with two disjoint inner contours into a skyrmion bag with a single inner contour containing a chiral kink (Fig.~\ref{fig:histogram_LD40}e).


Finally, two special SPs do not belong to any of the identified categories. They correspond to the progressive shrinking and eventual collapse of a chiral droplet and an antiskyrmion into the FM background (see Fig.~\ref{fig:histogram_LD40}f).

A histogram of SP energies for all identified transitions, measured relative to the lower-energy endpoint of each transition, shows the distribution of energy barriers in the system (see Fig.~\ref{fig:histogram_LD40}f). 
Clearly, 
the energy barriers associated with the five identified mechanisms exhibit only a weak dependence on the specific host texture and are only marginally influenced by the presence of tails, chiral kinks, or closed contours. For example, nucleating a chiral kink on the outer contour of a skyrmion bag requires nearly the same energy as creating a second kink on a droplet to convert it into an antiskyrmion. This insensitivity to the underlying texture highlights the universal character of the identified SP categories.

A clear separation of energy scales of homotopies and non-homotopies is also observed for the chosen value of $L_D$, with homotopies exhibiting lower energy barriers than transitions that change the topological charge. Motivated by this observation, we next examine the universality of this hierarchy through a scaling analysis presented in the following section.

\subsection{Scaling analysis of transition mechanisms}
\label{sssec:scaling_isolated_stripe}
\begin{figure*}
    \centering
    \includegraphics{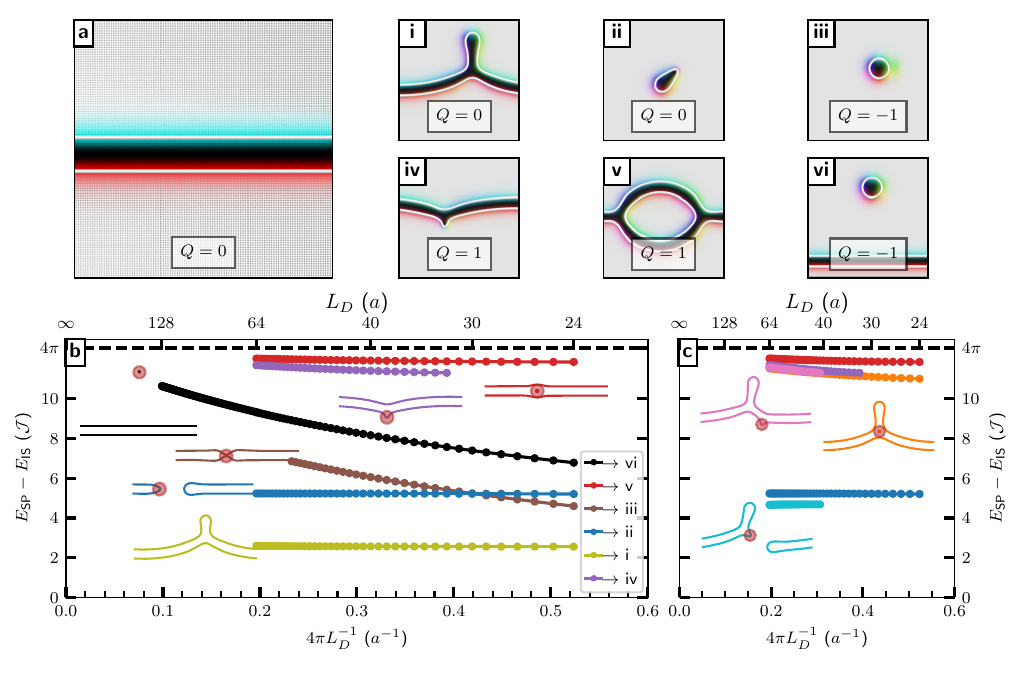}
    \caption{Isolated stripe state (\textbf{a}) and adjacent energy minimum states (\textbf{i}-\textbf{vi}). \textbf{b}: SP energies relative to the isolated stripe minimum as functions of $L_D$ at $h=0.623$. \textbf{c}: SP energies for transformations of an isolated stripe with an attached tail as functions of $L_D$ at $h=0.623$. Insets show SP configurations represented by the $m_z=0$ contours. Semi-transparent red dots indicate the position of the maximum of $|\chi|$ in SP configurations associated with transformations with $\Delta Q\ne0$.
    %
    %
    }
    \label{fig:vary_LD_stripe}
\end{figure*}

To further assess the universality of the identified transition mechanisms, we consider a simple host texture: a straight isolated stripe (Fig.~\ref{fig:vary_LD_stripe}a), which can also be seen as a fragment of an infinitely large skyrmion bag. The five SP categories —tail formation, chiral kink nucleation, contour creation, contour merging, and contour merging with chiral kink nucleation — are analyzed in this context.

Figure~\ref{fig:vary_LD_stripe}b shows the corresponding energy barriers as a function of $L_D$ ranging from small values, where neighboring magnetic moments exhibit strong canting, to large values, where the canting is small in the absence of discontinuities in the texture, and the lattice model approaches the continuum limit. In this limit, the energy barrier for nucleating a skyrmion in the ferromagnetic background (Fig.~\ref{fig:vary_LD_stripe}vi) converges to $4\pi \mathcal{J}$~\cite{heil2019}. Energies of all considered SPs in the isolated stripe system remain below this limit for all considered values of $L_D$.

There is a clear distinction in how homotopic and non-homotopic transitions behave under variation of $L_D$. The energy barriers associated with homotopies -- tail formation (Fig.~\ref{fig:vary_LD_stripe}i) and homotopic contour merging (Fig.~\ref{fig:vary_LD_stripe}ii) -- remain nearly independent of $L_D$. In contrast, the barriers for non-homotopic mechanisms -- non-homotopic contour merging (Fig.~\ref{fig:vary_LD_stripe}iii), chiral kink nucleation (Fig.~\ref{fig:vary_LD_stripe}iv), and contour creation (Figs.~\ref{fig:vary_LD_stripe}v,vi) -- increase with $L_D$ and appear to approach finite values in the continuum limit.

A crossover between the homotopic and non-homotopic contour-merging mechanisms occurs at $L_D \approx 30a$. For lower $L_D$, the non-homotopic transition becomes energetically favorable. The fact that the homotopy does not always yield the lowest energy barrier raises the question of whether an MEP representing a homotopy always exists between states with the same topological charge. This question is addressed in the following section.

Finally, Fig.~\ref{fig:vary_LD_stripe}c shows that the presence of a tail on the isolated stripe has only a minor influence on the energy barriers associated with homotopic contour merging, chiral kink nucleation, and contour creation (Fig.~\ref{fig:vary_LD_stripe}c). 
This insensitivity to local features of the host texture is further confirmed in Appendix~\ref{sec:app_curved_lines}, where calculations applied to a curved isolated stripe show that the corresponding energy barriers depend only weakly on the curvature of the host contour. 

\subsection{Homotopies and MEPs}
\label{ssec:existenceMEP_homotopy}

There exists an infinite number of homotopies between textures with the same topological charge. In this section, we investigate under what conditions an MEP can represent a homotopy. Although homotopy is only an approximate notion in the discrete lattice model considered here, it remains a useful framework for analyzing transitions. We focus on the collapse of the skyrmion bag state ($Q=0$) into the FM state for $L_D=30a$.

\begin{figure*}
    \centering
    \includegraphics{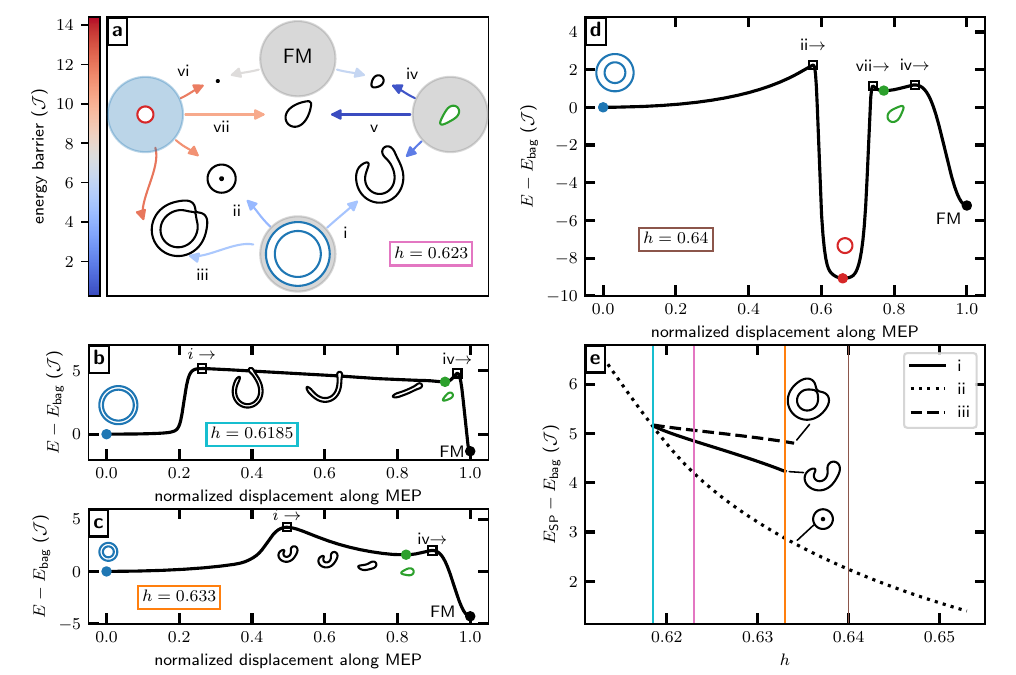}
    \caption{\textbf{a}: Part of the network of metastable states in a chiral magnet at $h=0.623$ and $L_D=30a$. Energy-minimum and SP configurations are represented by their $m_z=0$ contours. The arrows pointing from the adjacent minima toward the SPs are colored according to the corresponding energy barriers. \textbf{b-d}: Energy profiles along MEPs connecting the skyrmion bag state to the FM state at $h=0.6185$, $0.633$, and $0.64$, respectively. Selected configurations along the MEPs are shown by their $m_z=0$ contours. \textbf{e}: Energy barriers for the transitions \textbf{i-iii} as functions of $h$. Vertical lines indicate the field values used in \textbf{a-d}.
 }
    \label{fig:homotopy}
\end{figure*}

A homotopy-MEP is illustrated in Fig.~\ref{fig:homotopy}a. The transformation begins with homotopic contour merging accompanied by the nucleation of a chiral kink (i), converting the skyrmion bag into a chiral droplet. The droplet then continuously collapses into the FM state without forming singularities (iv).

Other collapse mechanisms are possible, but they do not represent homotopies. The skyrmion bag may first transform into a skyrmion either by collapse of the inner contour (ii) or via non-homotopic contour merging (iii). From the skyrmion state, the system can reach the FM state either directly (vi) or via the chiral droplet state (vii).

The homotopy-MEP, however, exists only for magnetic fields in the range $0.6185\le h\le 0.633$. The disappearance of the homotopy-MEP for the fields outside this range can be understood by analyzing the energy profiles at the critical fields (Figs.~\ref{fig:homotopy}b–c). Near the lower critical field, 
$h_l\approx0.6185$, the homotopic contour merging is followed by a contraction of the resulting structure toward the chiral droplet; however, this contraction is accompanied by only a weak decrease in energy. 
Below this field, the contraction becomes energetically unfavorable, and the configuration instead elongates toward an extended stripe domain.

With increasing field, the skyrmion bag becomes progressively more compact, enhancing the tendency toward collapse of the inner contour (Figs.~\ref{fig:homotopy}c). A slight increase of the external field beyond $h_u\approx0.633$ destabilizes the homotopy-MEP: the GMMF method converges to the SP corresponding to inner-contour collapse (ii) even when initialized from the homotopic contour-merging SP obtained at $h=0.633$. As a result, the actual MEP passes through the skyrmion state and does not represent a homotopy. The corresponding energy variation is shown in Fig.~\ref{fig:homotopy}d.

Interestingly, the homotopic contour-merging mechanism never provides the lowest energy barrier for the collapse of the skyrmion bag. The inner-contour collapse (ii), which is a non-homotopy, corresponds to a lower barrier, while the non-homotopic contour merging (iii) yields a larger barrier, as summarized in Fig.~\ref{fig:homotopy}be.

\section{Conclusions and discussion}
\label{sec:conclusion}
The SPSF developed in this work offers a systematic method to identify SPs on the system's energy surface that surround a given initial energy minimum corresponding to a metastable state of a magnetic system. 
The framework is particularly suitable for systems capable of hosting various co-existing localized textures. 
The method extends existing SP search algorithms~\cite{muller2018,bocquet2023,schrautzer2025} by introducing a structured multi-stage design -- comprising preprocessing, escape, convergence, and postprocessing stages -- that enables a systematic and automated exploration of the energy surface around minima. 
A key methodological advancement is the escape stage, where deformations of the energy-minimum configuration are sampled along low-energy excitations of symmetry-informed subsystems. 
Each escape attempt generates a tendency toward an SP by locally breaking the symmetry of the initial configuration. Sampling low-energy excitations of subsystems centered within the fundamental domain makes the procedure systematic and computationally efficient.
Combination of the escape stage, which requires calculations only within a small subset of the system, with the efficient implementation of the GMMF method for convergence to SPs~\cite{schrautzer2025} makes the methodology applicable to large systems and to systems with long-range interactions. 
For localized magnetic textures in two-dimensional chiral magnets, the SPSF reveals a complex network of metastable states connected by first-order SPs. Despite the complexity, the metastable textures can be described using three basic structural elements: closed contours, chiral kinks~\cite{kuchkin2020}, and tails~\cite{kuchkin2023}.

All identified SPs correspond to five categories: chiral-kink nucleation, contour merging, contour creation through injection of a Bloch-point-like defect, tail formation, and contour merging accompanied by simultaneous chiral-kink nucleation. The last two mechanisms correspond to homotopies, whereas the remaining ones change the topological charge of the system. The energy barriers associated with these mechanisms depend only weakly on the specific host texture, which suggests that the SP categories have a universal character. 

The scaling behavior of the transition mechanisms further distinguishes homotopies and non-homotopies. Barriers for transformations that change the topological charge increase as the continuum limit is approached, whereas barriers for homotopies remain nearly independent of $L_D$. At the same time, all barriers remain below the known nucleation barrier of an isolated skyrmion in the FM background in the continuum limit~\cite{heil2019}.

Although any two textures with the same topological charge $Q$ can always be continuously transformed into one another without changing the total topological charge of the system, our results show that, under certain conditions, no homotopy between the states corresponds to an MEP. In such cases, the MEP necessarily involves the nucleation and subsequent annihilation of topological charge. Even if MEPs describing homotopies exist, they do not necessarily correspond to the lowest energy barrier between the states. 
%

The idea of sampling symmetry-informed subsystem excitations can be straightforwardly applied to a broad range of magnetic materials including three-dimensional systems.
Further improvements of the SPSF are possible. 
On the one hand, techniques such as genetic multi-objective algorithms~\cite{xu2026} could be used to learn the optimal distributions of subsystems yielding the richest diversity of SPs.
On the other hand, the number of convergence stage calculations reentering a convex region could be further reduced by implementing re-escaping strategies~\cite{gunde2024}.

Due to its modular structure and efficiency, the SPSF is ideally suited to be used in advanced methods that rely on repeated SP searches. 
For example, the energy surface can be explored globally by searching for metastable states via recursive traversing from one energy minimum to another over SPs. 
Combining this with on-the-fly calculations of the transition rates  at a given temperature using HTST~\cite{bessarab2012} or Kramers/Langer theory~\cite{fiedler2012,bocquet2024} yields the adaptive kinetic Monte Carlo method~\cite{henkelman2001,jonsson2011} for simulating the long timescale dynamics of a system.

SP searches can also serve as a foundation for global optimization of an objective function in a wider context~\cite{plasencia2014}, as well as for path optimization~\cite{einarsdottir2012}, such as in the calculation of tunneling within instanton theory~\cite{asgeirsson2018,vlasov2020} and the propagation of radio waves~\cite{nosikov2020}. 


\begin{acknowledgments}
The authors would like to thank S. Heinze, M. Sallermann, and T. Sigurj\'onsd\'ottir for fruitful discussions. This work was funded by the Icelandic Research Fund (Grants No. 239435, No. 2410333, and No. 217750), the University of Iceland Research Fund (Grant No. 15673), the Swedish Research Council (Grant No. 2020-05110), and the Crafoord Foundation (Grant No. 20231063). The calculations were carried out at the high-performance computing resources available at the Kiel University Computing Centre and at the Icelandic Research e-Infrastructure facility supported by the Icelandic Infrastructure Fund.
%
%
%
%
\end{acknowledgments}

\appendix

\section{Correspondence between the lattice and continuum descriptions of magnetization}
\label{app:continuous_discrete}
We start from the square-lattice Hamiltonian [see Eq.~(\ref{eq:hamiltonian})], equipped with first- and third-neighbor couplings $J_1$, $J_3$, $D_1$, and $D_3$, while setting $J_2=D_2=0$. 
After Fourier transforming the magnetic moments and expanding the resulting trigonometric functions up to fourth order in the wave vector, one obtains the following micromagnetic energy functional~\cite{heil2019,derras2019}:
%
\begin{align}
\mathcal{E}=\mathcal{E}_{\text{Zee}}+\mathcal{E}_\text{ex}^{\text{FM}}+\mathcal{E}_\text{ex}^{(2)}+\mathcal{E}_\text{ex}^{(4)}+\mathcal{E}_\text{DM}^{(1)}+\mathcal{E}_\text{DM}^{(3)}~,
\label{eq:toward_continuum_ham}
\end{align}
where the individual terms are given by:
\begin{align}
\mathcal{E}_{\text{Zee}}=&-\int M_S\vec{B}\cdot\vec{m}(\vec{r})\dd{r}^2,\\
    \mathcal{E}_\text{ex}^{\text{FM}}=&-\frac{2(J_1+J_3)}{a^2}\int\dd{r}^2,\\
\mathcal{E}_{\text{ex}}^{(2)}= &\int \frac{J_1+4J_3}{2}\sum_{\alpha=x,y}(\partial_\alpha\vec{m})^2\dd{r}^2,\\
     \mathcal{E}_{\text{ex}}^{(4)}= &-\int a^2\frac{J_1+16J_3}{24}\sum_{\alpha=x,y}(\partial_\alpha^2\vec{m})^2\dd{r}^2\label{eq:ex_order_4},\\
    \mathcal{E}_{\text{DM}}^{(1)}=&\int\frac{D_1+2D_3}{a}R_\beta\vec{m}\cdot\vec{\nabla}\times R_\beta\vec{m}\dd{r}^2,
\end{align}
\begin{widetext}
\begin{equation}
    \mathcal{E}_\text{DM}^{(3)}=\int\frac{a(D_1+8D_3)}{6}R_\beta\vec{m}(\vec{r})\cdot\left[\sum\limits_{\alpha=x,y}\hat{\alpha} \times \partial_{\alpha}^3(R_\beta\vec{m})\right]\dd{r}^2\label{eq:dm_order_3}.
\end{equation}
\end{widetext}
The parameters $J_1$, $J_3$, $D_1$, and $D_3$ of the lattice model are chosen such that the leading higher-order corrections, $\mathcal{E}_\text{ex}^{(4)}$ and $\mathcal{E}_\text{DM}^{(3)}$, vanish (see Table~\ref{tab:modelparameters}).
%

\section{Influence of escape-stage thresholds on the outcome of the convergence stage}
\label{sec:app_escape}

The escape stage is characterized by two numerical thresholds. The threshold
$\lambda_\text{esc}$ controls how far the configuration must move beyond the convex region of the initial energy minimum, as measured by the lowest Hessian eigenvalue [Eq.~\eqref{eq:escape_crit_single}]. The threshold
$w_\text{esc}$ controls the alignment between the energy gradient and the minimum mode [Eq.~\eqref{eq:valley_entrance}], which is used as an indicator that the system has entered an ascending valley of the energy surface~\cite{hoffman1986,jorgensen1988,schlegel1992,bofill2012}.

The influence of these thresholds is assessed by applying the SPSF to the skyrmion state. The setup is the same as in Sec.~\ref{ssec:results_methodpresentation}, giving $180$ escape attempts. For each choice of the threshold values, all $180$ attempts are successful and provide starting configurations for the convergence stage. Each convergence-stage calculation then either reaches a first-order SP or re-enters a convex region.

\begin{figure}
    \centering
    \includegraphics[width=1.0\linewidth]{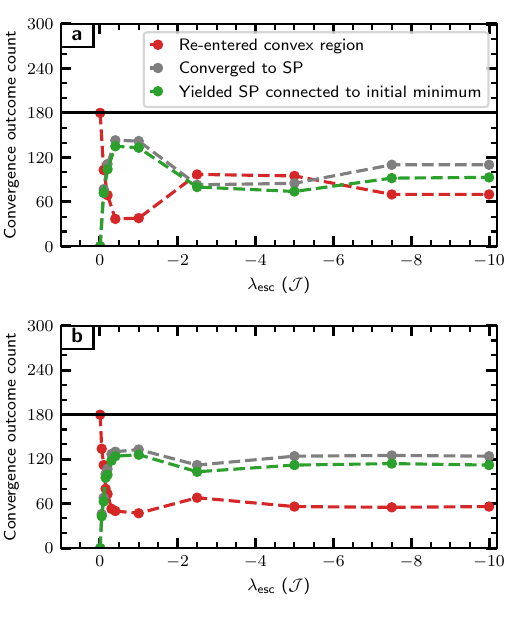}
    \caption{Outcome of the convergence stage for the $180$ escape attempts from the skyrmion state at $h=0.623$ and $L_D=40a$, as a function of the escape threshold $\lambda_{\rm esc}$. The horizontal black line indicates the total number of escape attempts. Red, gray, and green curves show the numbers of convergence attempts ending in a convex region, reaching a first-order SP, and reaching a first-order SP connected to the initial skyrmion-state minimum, respectively. \textbf{a}: Results obtained using only the eigenvalue-based escape criterion [Eq.~\eqref{eq:escape_crit_single}]. \textbf{b}: Results obtained when the alignment criterion is also included, with $w_{\rm esc}=0.25$.
    %
    }
    \label{fig:res_sk_escvar}
\end{figure}

Figure~\ref{fig:res_sk_escvar}a shows the outcome when only the eigenvalue criterion in Eq.~\eqref{eq:escape_crit_single} is used. For values of $\lambda_\text{esc}$ close to zero, most convergence-stage calculations re-enter a convex region, and only few SPs are obtained. Making $\lambda_\text{esc}$ more negative initially improves the outcome, and the number of successful convergence attempts reaches a maximum around $\lambda_\text{esc}\approx -0.2\mathcal{J}$. Beyond this point, further decreasing $\lambda_\text{esc}$, however, increases the number of re-entrances. Thus, when only the $\lambda_\text{esc}$ criterion is used, the outcome of the convergence stage is rather sensitive to the chosen threshold value, whose optimum is not known \textit{a priori} and may vary with the initial state and model parameters. 

This sensitivity is strongly reduced when the alignment criterion is included. As shown in Fig.~\ref{fig:res_sk_escvar}b for $w_\text{esc}=0.25$, the outcome of the convergence stage becomes nearly independent of $\lambda_\text{esc}$ once $\lambda_\text{esc}\lesssim -0.1\mathcal{J}$. In this regime, about two thirds of the convergence-stage calculations reach first-order SPs, whereas about one third re-enters a convex region. Most of the identified SPs are connected to the initial skyrmion state.

\begin{figure}
    \centering
    \includegraphics[width=1.0\linewidth]{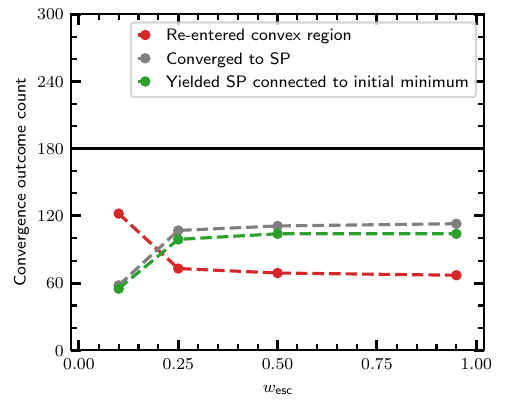}
    \caption{Outcome of the convergence stage for the $180$ escape attempts from the skyrmion state at $h=0.623$ and $L_D=40a$, as a function of the alignment threshold $w_{\rm esc}$ at fixed $\lambda_{\rm esc}=-0.2\mathcal J$. The horizontal black line indicates the total number of escape attempts. Red, gray, and green curves show the numbers of convergence attempts ending in a convex region, reaching a first-order SP, and reaching a first-order SP connected to the initial skyrmion-state minimum, respectively.
%
}
    \label{fig:skyrmion_esc_w_var}
\end{figure}

Finally, Fig.~\ref{fig:skyrmion_esc_w_var} shows how the outcome of the convergence stage depends on $w_\text{esc}$ at fixed $\lambda_\text{esc}=-0.2\mathcal{J}$. For $w_\text{esc}\leq 0.1$, most of the convergence attempts end in a convex region. However, the number of successful convergence attempts increases with increasing $w_\text{esc}$ and reaches a plateau at $w_\text{esc}\approx 0.25$, where about two thirds of the calculations reach SPs. These results motivate the use of the alignment criterion with a moderate threshold $w_\text{esc}$, which makes the escape stage more robust with respect to the choice of $\lambda_\text{esc}$.

\section{Effect of contour curvature on energy barriers}
\label{sec:app_curved_lines}
To test whether the energy barriers depend on the curvature of the host contour, we consider a curved isolated stripe at $L_D=48a$ and $h=0.623$. The stripe is embedded in a system of $200\times 200$ magnetic moments. The outer eight rows and columns of magnetic moments are kept fixed to maintain the curved-stripe configuration. 

\begin{figure}
    \centering
    \includegraphics{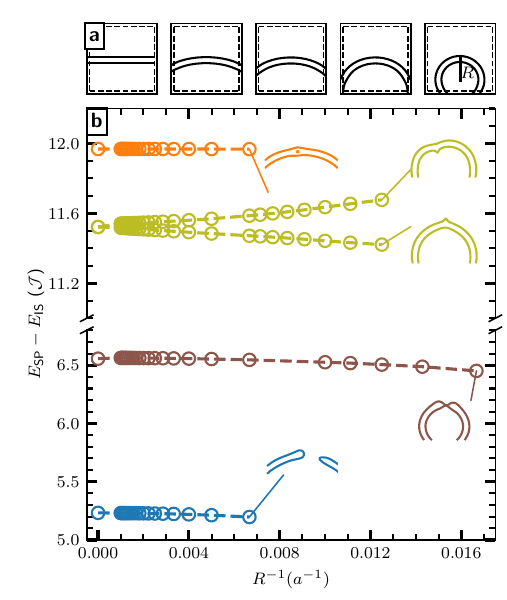}
    \caption{Effect of contour curvature on the energy barriers of transformations involving an isolated stripe at $L_D=48a$ and $h=0.623$. \textbf{a}: Isolated stripes with different radii $R$, shown by the $m_z=0$ contours. The dashed and solid lines indicate the boundaries of the region in which the magnetic moments are kept fixed to preserve the imposed shape of the stripe. \textbf{b}: SP energies relative to the isolated-stripe minimum, $E_{\rm SP}-E_{\rm IS}$, as functions of the inverse radius of the stripe. Representative SP configurations are shown by their $m_z=0$ contours.
%
%
%
}
    \label{fig:curved_isolated_stripes}
\end{figure}

Figure~\ref{fig:curved_isolated_stripes}a shows the reference configurations. The contour curvature is varied by changing the radius $R$ of the curved stripe. The obtained SPs correspond to the same transition mechanisms as for the straight isolated stripe in Sec.~\ref{sssec:scaling_isolated_stripe}: creation of a contour via a Bloch-point-like defect, nucleation of a chiral kink, homotopic contour merging, and non-homotopic contour merging.

Fig.~\ref{fig:curved_isolated_stripes}b demonstrates that, for the range of curvatures considered here, the barriers associated with these transition mechanisms are not strongly affected by the curvature of the host contour. This supports the conclusion of Sec.~\ref{sssec:results_graph} that the energy barriers are primarily determined by the transition mechanism rather than by the detailed shape of the host texture.

\FloatBarrier

\makeatletter
\immediate\write\@auxout{\string\citation{apsrev42Control}}
\makeatother

\bibliographystyle{apsrev4-2}
\bibliography{apssamp,apsrev42Control}
\end{document}